\documentclass{aa} 

\usepackage{graphicx}
\usepackage{amsmath,amssymb}
\usepackage{wasysym}
\usepackage{mathtools} 
\usepackage{booktabs} 
\usepackage[UKenglish]{isodate}

\usepackage{graphicx}

\usepackage{natbib,twoopt}

\usepackage{color}
\def\vsini{$v\sin i$}

\def\ha10{${\rm H}\alpha\,10\%$}

\begin{document} 

\title{Evidence of radius inflation in stars approaching the slow-rotator sequence}
\titlerunning{Radius inflation of stars approaching the slow-rotator sequence}

\author{
A.~C. Lanzafame\inst{\ref{inst1},\ref{inst2}} \and 
F. Spada\inst{\ref{inst3}}     \and
E. Distefano\inst{\ref{inst2}}
}

\authorrunning{A.~C. Lanzafame et al.}

\offprints{A.~C. Lanzafame \\ \email{a.lanzafame@unict.it}}

\institute{
Universit\`a di Catania, Dipartimento di Fisica e Astronomia, Sezione Astrofisica, Via S. Sofia 78, I-95123 Catania, Italy \label{inst1} \\
 \email{a.lanzafame@unict.it}
\and
INAF-Osservatorio Astrofisico di Catania, Via S. Sofia 78, I-95123 Catania, Italy\label{inst2}
\and
Leibniz-Institut f\"ur Astrophysik Potsdam (AIP), An der Sternwarte 16, D-14482, Potsdam, Germany\label{inst3}
}

\date{Received 2 May 2015 / Accepted 4 September 2016}

\abstract{
Average stellar radii in open clusters can be estimated from rotation periods and projected rotational velocities under the assumption that the spin axis has a random orientation.
These estimates are independent of distance, interstellar absorption, and models, but their validity can be limited by lacking data (truncation) or data that only represent upper or lower limits (censoring).
}{
We present a new statistical analysis method to estimate average stellar radii in the presence of censoring and truncation.
}{
We used theoretical distribution functions of the projected stellar radius $R \sin i$ to define a likelihood function in the presence of censoring and truncation.
Average stellar radii in magnitude bins were then obtained by a maximum likelihood parametric estimation procedure.
}{
This method is capable of recovering the average stellar radius within a few percent with as few as $\text{about ten}$ measurements. 
Here we apply this for the first time to the dataset available for the Pleiades. 
We find an agreement better than $\approx$ 10 percent between the observed $R$ vs $M_K$ relationship and current standard stellar models for $1.2\ge M/M_{\odot}\ge 0.85$ with no evident bias. 
Evidence of a systematic deviation at $2\sigma$ level are found for stars with $0.8\ge M/M_{\odot}\ge 0.6$ that approach the slow-rotator sequence.
Fast rotators ($P < 2$ d) agree with standard models within 15 percent with no systematic deviations in the whole $1.2 \apprge M/M_{\odot} \apprge 0.5$ range.
}
{
The evidence of a possible radius inflation just below the lower mass limit of the slow-rotator sequence indicates a possible connection with the transition from the fast- to the slow-rotator sequence.\thanks{Table 1 is only available in electronic form at the CDS via anonymous ftp to cdsarc.u-strasbg.fr (130.79.128.5)
or via http://cdsweb.u-strasbg.fr/cgi-bin/qcat?J/A+A/}
}


\maketitle


\section{Introduction}
\label{sec:Introduction}

The disagreement between theoretical and observed parameters of young magnetically active and of fully convective or almost fully convective low-mass stars remains one of the main long standing problems in stellar physics. 
Current investigations focus on the inhibition of the convective transport by interior dynamo-generated magnetic fields and/or by the blocking of flux at the surface by cool magnetic starspots 
\citep[e.g.][]{2001ApJ...559..353M, 
2007A&A...472L..17C, 
2013ApJ...779..183F,
2014ApJ...789...53F,
2014MNRAS.441.2111J},
which produce an increase in stellar radius and a decrease in $T_{\rm eff}$.
The same effect is also thought to be linked to the observed correlation between Li abundance and rotation \cite[e.g.][]{
2014ApJ...790...72S,
2015MNRAS.449.4131S,
2015ApJ...807..174S,
2014MNRAS.441.2111J}.
The consequences of these discrepancies are manifold.
These include, for example, the determination of the mass and the radius of exoplanets, whose accuracy depends on that of the hosting star \citep[e.g.][]{2004ASPC..318..159H,2015ApJ...804...64M}, the age estimate of young open clusters \citep[e.g.][]{2014prpl.conf..219S,2015ApJ...807..174S}, and the mass-luminosity relationship for magnetically active low-mass stars.

Fundamental determinations of stellar masses and radii with a 3 percent accuracy or better are provided by the light-curve analysis of detached eclipsing binaries \citep[e.g.][]{2010A&ARv..18...67T,2012ApJ...757...42F}.
Interferometric angular diameter measurements of single stars are available today for tens of stars \citep[e.g.][]{2012ApJ...757..112B} with diameters measured to better than 5 percent.

Statistical methods based on the product of $P$ and \vsini, which produces the projected radius $R \sin i$ (Sect.\,\ref{sec:ProjectedRadius}), and the assumption of random orientation of the spin axis \citep[e.g.][]{2009MNRAS.399L..89J} have the advantage of providing mean radii estimates for a large number of (coeval) single stars independently of distance, interstellar absorption, and models.
No evidence of preferred orientation of the spin axis in open clusters has been found so far \citep[e.g.][and references therein]{2010MNRAS.402.1380J}, and therefore the method seems to be sound in this respect.
The main difficulty is that the data sample is always truncated at a combination of sufficiently low inclination angle $i$ and low equatorial velocities $v_{\rm eq}$.
In these cases, depending also on the spectral resolution, \vsini\ cannot be derived and only an upper limit can be given.
A low $i$ may also cause difficulties in measuring $P$ and therefore there may be cases in which either one or both \vsini\ and $P$ cannot be measured.
At the other extreme, ultra-fast rotator spectra can be so smeared by the rotational broadening that in some cases only a lower \vsini\ limit can be given.

To take a low $R \sin i$ truncation into account, \cite{2009MNRAS.399L..89J} considered a cut-off inclination such that stars with lower inclination yield no $R \sin i,$ and they corrected the average $\sin i$ accordingly.
Mean radii are then derived by taking the average of the ratio $R \sin i / \left< \sin i \right> $ in suitable magnitude bins.

Here we present a new method, based on the survival analysis concept \citep{Klein+2003}, that makes use of the whole information content of the dataset by also considering upper and lower limits and data truncation.
Data may also come from inhomogeneous estimates, like those in which \vsini\ upper and lower limits are obtained from different analyses and instrumentation, as long as they are not affected by significant biases.
Uncertainties due to surface differential rotation (SDR) are also estimated, with the most likely values derived from the recent work of \cite{Distefano_etal:2016}.
The method is applied for the first time to the rich dataset available for the Pleiades. 

In Sect.\,\ref{sec:data} we present the data used in this work. The method is described in Sect.\,\ref{sec:Method}. 
The results obtained for the Pleiades dataset are discussed in Sect.\,\ref{sec:Pleiades}. 
We draw our conclusions in Sect.\,\ref{sec:Conclusions}.

\section{Data}
\label{sec:data}

For this work rotational periods and memberships from \cite{Hartman_etal:2010} are used.
Measurements of \vsini\ are taken from \cite{1987ApJ...318..337S}, \cite{1993ApJS...85..315S}, \cite{1998A&A...335..183Q} and \cite{2000AJ....119.1303T}. 
Magnitudes are adopted from \cite{Stauffer_etal:2007}.

\begin{figure}[ht]
\begin{center}
\includegraphics[width=0.5\textwidth]{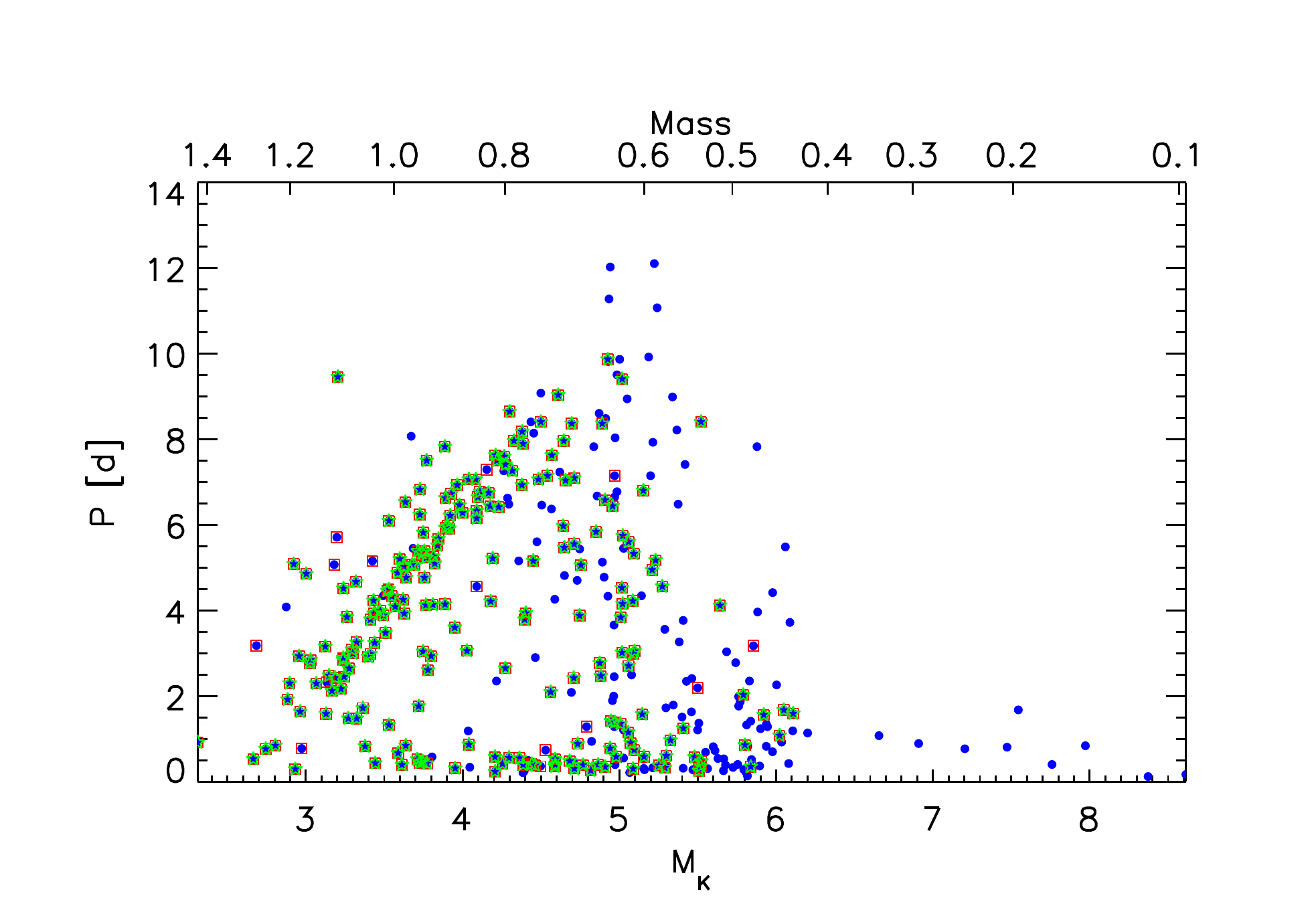}
\includegraphics[width=0.5\textwidth]{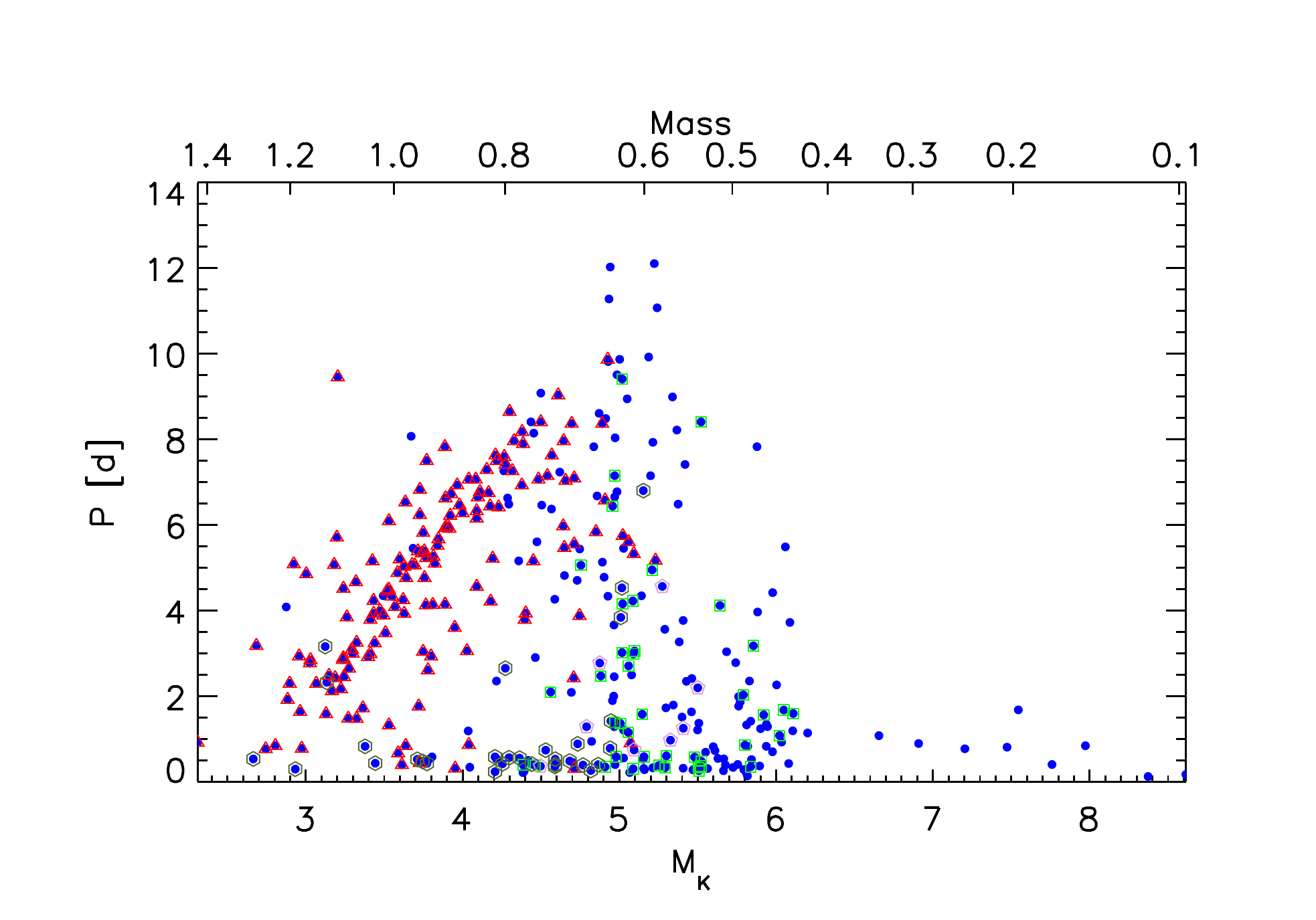}
\caption{\cite{Hartman_etal:2010} $P$ dataset of the Pleiades (blue filled circles). 
Upper panel: open red squares outline targets for which \vsini\ is available and green stars those considered in this work. 
Lower panel: Open red triangles indicate \vsini\ from \cite{1998A&A...335..183Q}, 
green squares the values adopted from \cite{2000AJ....119.1303T}, 
plum pentagons those from  \cite{1987ApJ...318..337S}, 
and olive hexagons those from \cite{1993ApJS...85..315S}.
See text for details.
}
\label{fig:data}
\end{center}
\end{figure}

The  $P$ dataset of \cite{Hartman_etal:2010} comprises 383 stars and is 93 percent complete in the mass range $1.0 \apprge M/M_{\odot} \apprge 0.7$.  
Of these, 227 have measured \vsini.
Stars flagged as binaries in \cite{Hartman_etal:2010} were excluded from the sample.
A total of 217 stars constitute the final working sample (Fig.\,\ref{fig:data}).

Theoretical mass-radius-magnitude relationships were taken from \cite{2015A&A...577A..42B}.
Together with the adopted distance to the Pleiades of 136.2 pc \citep{2014Sci...345.1029M}, an age of 120 Myr \citep{Stauffer_etal:1998}, and an extinction of $A_K = 0.01$ mag \citep{Stauffer_etal:2007}, the models of \cite{2015A&A...577A..42B} were used to build bins in $M_K$ magnitude corresponding to approximately regular intervals of mass, and then in the comparison of our results with the theoretical $R$-$M_K$ relationship.
The calculations reported in Sect.\,\ref{sec:Pleiades} were repeated, also assuming a distance to the Pleiades of 120.2 pc \citep{2009A&A...497..209V}.
Our results, however, are more consistent with a distance of $\approx 136.2$ pc, and therefore we report only the results obtained assuming this value \citep[see also][]{2005AJ....129.1616S}.

We note that the fraction of stars with both $P$ and \vsini\ considered in the analysis with respect to the whole $P$ dataset of \cite{Hartman_etal:2010} is very close to one down to $M_K \approx 4.3$, corresponding to $M \approx 0.8 M_{\odot}$.
For fainter magnitudes this fraction decreases progressively to $M_K \approx 6$ ($M \approx 0.5 M_{\odot}$), below which there are no \vsini\ measurements and very few and sparse $P$ measurements (Fig.\,\ref{fig:data}, upper panel).
Furthermore, \vsini\ values for fast rotators were mostly adopted from \cite{1993ApJS...85..315S}, for slow-rotators with $M_K \apprle 4.3$ mostly from \cite{1998A&A...335..183Q}, while at fainter magnitudes the measurements are mostly those reported by \cite{2000AJ....119.1303T}.
Possible consequences of this inhomogeneity for our results are discussed in Sect.\,\ref{sec:Pleiades}.

\begin{table*}
   \centering
      \caption{Data with the rotational sequence classification based on the $P$ vs $(B-V)$ 
      diagram adapted from \cite{2015A&A...584A..30L}.
      The full table is available in electronic form at CDS and contains 217 stars.
      Mass and radius are derived from $M_K$ according to the models of \cite{2015A&A...577A..42B}.
      Periods are truncated to the 6$^{th}$ decimal figure.
      Upper \vsini\ limits are flagged with ${\rm lim}_{v \sin i}=1$, lower limits with ${\rm lim}_{v \sin i}=2.$
      See text for details.}
      \label{tab:Pdata}
      \begin{tabular}{cccccccccrc}
      \hline
  ID             & RA (deg)  & DEC (deg)  & $(B-V)$ & $M_K$ & $M$ & $R$ &  $P$   & ${\rm lim}_{v \sin i}$ & \vsini  & seq. \\
                 & J2000     & J2000      & (mag)   & (mag) & $M_{\odot}$ & $R_{\odot}$ & (d) &           & km s$^{-1}$ & \\
\hline
  HAT214-0001101 & 52.890079 & 26.265511  & 0.68 & 3.387 & 1.050 & 0.961 &  3.242160 &  0  & 15.5 &  i \\
  HAT259-0005281 & 53.001961 & 23.774900  & 1.32 & 4.839 & 0.664 & 0.596 &  8.366860 &  0  &  5.5 &  - \\
  HAT259-0001868 & 53.307941 & 23.006470  & 0.95 & 3.988 & 0.876 & 0.778 &  7.064630 &  0  &  3.1 &  i \\
  HAT259-0000955 & 53.507519 & 24.880960  & 0.70 & 3.497 & 1.017 & 0.924 &  4.333200 &  0  & 11.1 &  i \\
  HAT259-0000962 & 54.073441 & 21.894220  & 0.69 & 3.569 & 0.996 & 0.900 &  4.252010 &  0  &  8.5 &  i \\
  HAT259-0002206 & 54.126259 & 24.012230  & 0.97 & 4.222 & 0.812 & 0.719 &  7.410650 &  0  &  4.9 &  i \\
  HAT259-0002463 & 54.301289 & 21.468170  & 1.04 & 4.265 & 0.801 & 0.709 &  7.262440 &  0  &  5.5 &  - \\
  HAT259-0000543 & 54.594090 & 22.499701  & 0.57 & 3.015 & 1.162 & 1.105 &  2.295330 &  0  & 24.4 &  i \\
  HAT259-0000690 & 54.736938 & 24.569799  & 0.65 & 3.360 & 1.058 & 0.971 &  2.986200 &  0  &  9.6 &  i \\
  HAT259-0000652 & 54.806122 & 24.466511  & 0.62 & 3.224 & 1.099 & 1.021 &  2.646660 &  0  & 11.8 &  i \\
\hline
      \end{tabular}
\end{table*}

\section{Method}
\label{sec:Method}

\subsection{Projected radius}
\label{sec:ProjectedRadius}

When we assume spherical symmetry, the relationship between stellar radius $R$, stellar equatorial rotational period $P_{\rm eq}$ , and equatorial velocity $v_{\rm eq}$ is
\begin{equation}
\label{eq:Radius}
R = \mathcal{C} P_{\rm eq} v_{\rm eq} 
,\end{equation}
where 
$\mathcal{C} =0.02$ when $P_{\rm eq}$ is in days, $R$ in solar units, and $v_{\rm eq}$ in km s$^{-1}$.
The analysis of photometric time-series of stars showing rotational modulation produces the measured period $P$ (see Sect.\,\ref{sec:SDR}), while the estimated rotational broadening from spectroscopic analysis provides projected rotational velocities \vsini. 
For most of our sample deviations from spherical symmetry are expected to be negligible \citep[e.g.][]{1995ApJ...439..860C}.
Furthermore, from our SDR estimate (Sect.\,\ref{sec:SDR}) and the work of \cite{2003A&A...408..707R}, \cite{2003A&A...412..813R}, and \cite{2010arXiv1010.5932V}, who also took limb darkening into account, we estimate that the SDR effects on \vsini\ are not larger than $\approx$1 percent and are therefore significantly smaller than our conservative estimate of the period uncertainties associated with SDR (see Sect.\,\ref{sec:SDR}).
For our purposes we therefore neglected the SDR effects on \vsini\ and adopted the relationship
\begin{equation}
(v\sin i) = v_{\rm eq} \sin i ,
\end{equation}
where $i$ is the inclination of the spin axis from the line of sight.
Combining the measured \vsini\ and $P_{\rm eq}$ , we obtain the projected stellar radius
\begin{equation}
\label{eq:ProjectedRadius}
(R \sin i) = \mathcal{C} P_{\rm eq} v_{\rm eq} \sin i = \mathcal{C} P_{\rm eq} ~ (v \sin i).
\end{equation}

The inclination angle $i$ is unknown and therefore it is not possible to derive either $v_{\rm eq}$ from \vsini\ or $R$ from $R \sin i$ for each individual star.
However, when the underlying probability density functions of $R$ and $i$ are known, it is possible to estimate the expected value of $R$, $\mathcal{R}$, from an ensemble of $R \sin i$ measurements (see Sect.\,\ref{sec:DistributionFunctions}).

\subsection{Surface differential rotation}
\label{sec:SDR}

Surface differential rotation implies that magnetically active regions, associated with dark spots and bright faculae, rotate with different frequencies, depending on their latitudes.
Multiple active regions at different latitudes broaden the periodogram peak and contribute to the uncertainty in the period, as discussed by \cite{Hartman_etal:2010}, for instance.
On the other hand, SDR also leads to a systematic error in determining the rotation period of the star since the degree of differential rotation and the latitude of the dominant active region are not known, which prevents us from relating the measured period to the equatorial period $P_{\rm eq}$.
For young rapidly rotating stars like the Pleiades, \cite{Hartman_etal:2010} assumed that the dominant active region groups can be assumed to be isotropically distributed, from which they estimated that the mean rotational period is $\left<P\right> = 1.03 P_{\rm eq}$.
Following \cite{2005PhyU...48..449K}, for a solar-like SDR\footnote{It is customary to indicate an SDR corresponding to a decrease in surface rotational velocity with latitude as a solar-like SDR.} in which spots are confined to latitudes $|\beta| < 30^{\circ}$, \cite{Hartman_etal:2010} estimated $\left<P\right> = 1.07 P_{\rm eq}$.

\begin{figure}[ht]
\begin{center}
\includegraphics[width=0.44\textwidth,angle=270,trim={0 2cm 0 0},clip]{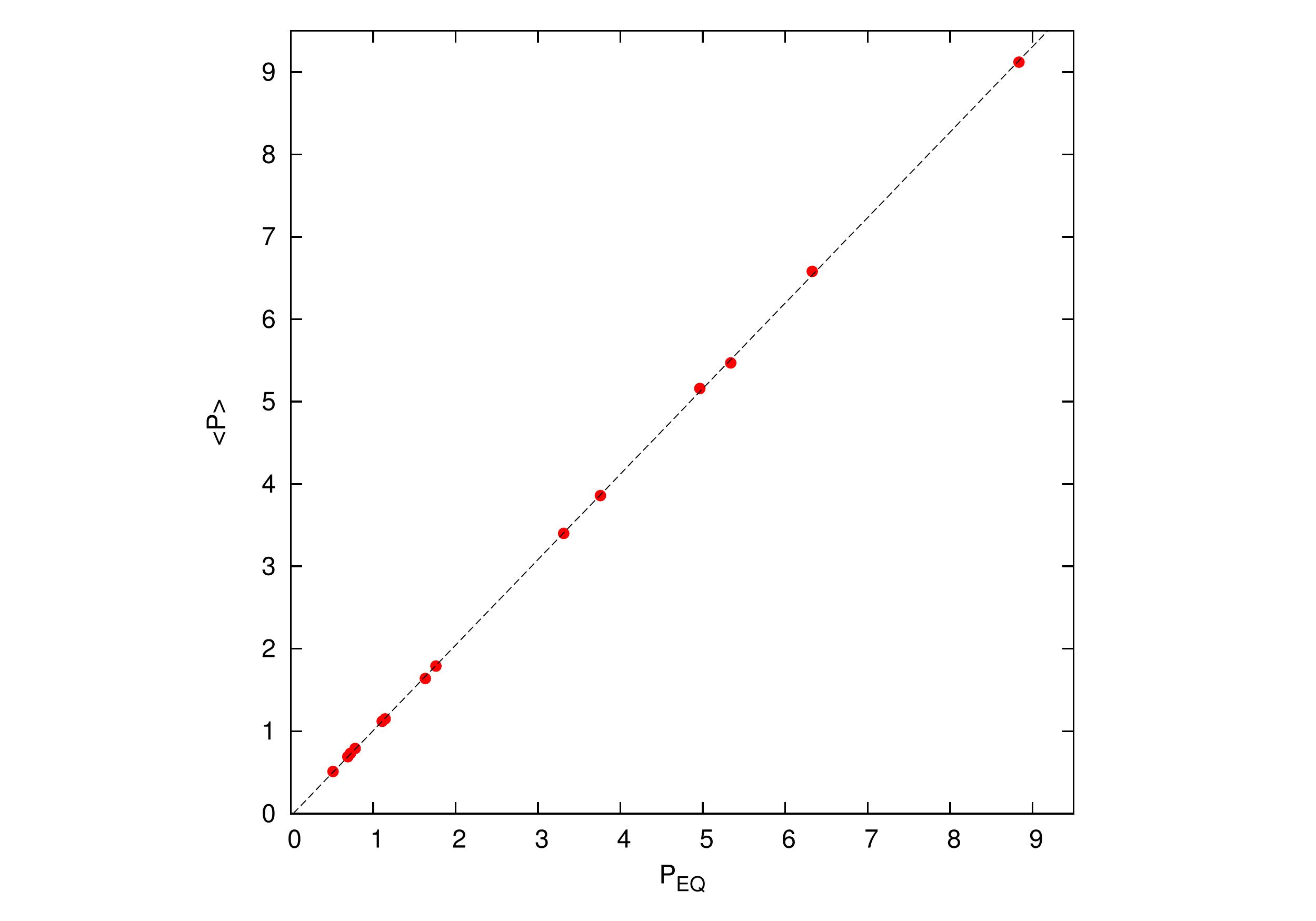}
\caption{Fit to the mean period vs equatorial period for the AB\,Dor young loose association.
}
\label{fig:ABDorPeriodFit}
\end{center}
\end{figure}

\cite{Distefano_etal:2016} made use of long-term photometric monitoring to estimate SDR lower limits in members of young loose association from the $P$ modulation itself.
Considering the AB\,Doradus young loose association, whose age is very similar if not identical to the Pleiades, assuming a solar-like SDR and that the minimum period corresponds to $P_{\rm eq}$, we fitted the observed $\left<P\right>$ vs $P_{\rm eq}$ with a linear relationship obtaining $\left<P\right> = 1.04 P_{\rm eq}$ (Fig.\,\ref{fig:ABDorPeriodFit}), which is very similar to that estimated by \cite{Hartman_etal:2010}.

We therefore assumed that the \cite{Hartman_etal:2010} $P$ are representative of $\left<P\right>$ and derived the equatorial period from the $\left<P\right> = 1.04 P_{\rm eq}$ relationship.
In this way, we obtained our best estimate of $P_{\rm eq}$.
Furthermore, to evaluate the uncertainties associated with the SDR,
we considered at one extreme solid-body rotation, $\left<P\right> = P_{\rm eq}$, and at the other extreme the $\left<P\right> = 1.07 P_{\rm eq}$ relationship. 
Given the information derived from the AB\,Dor mean vs equatorial period fit, such uncertainties likely overestimate the true uncertainties due to the SDR.
The corresponding $R \sin i$ uncertainties obtained using Eq.\,\eqref{eq:ProjectedRadius} are then also overestimated and larger, overall, than expected from the propagation of $P_{\rm eq}$ and \vsini\ uncertainties due to SDR (see Sect.\,\ref{sec:ProjectedRadius}). 
These can therefore be considered to include both $P_{\rm eq}$ and \vsini\ uncertainties due to the SDR.

\subsection{Distribution functions}
\label{sec:DistributionFunctions}

The statistical analysis presented here is similar to that used extensively in the past to derive the expected distribution of $v_{\rm eq}$  from \vsini\ measurements \citep[e.g.][]{1993A&A...269..267G,1998A&A...335..183Q,2011MNRAS.416..447S}.
The application of these concepts to the analysis of the $R \sin i$ distribution in open clusters has some advantages with respect to the analysis of \vsini.
Stars of similar mass in a stellar cluster, which therefore have
approximately the same age, are expected to also have a similar radius, and this can justify the assumption  that radii are distributed normally around the expected value, with a standard deviation that includes the intrinsic spread and measurement uncertainties.
In contrast, there are no such constraints on $v_{\rm eq}$, which makes reconstructing its distribution conceptually more difficult. 
Following \cite{1950ApJ...111..142C}, the $y \equiv R \sin i$ probability density function (p.d.f.) for a random orientation of spin axis can be therefore written as 
\begin{equation}
\label{eq:pdf}
\phi(y | \mathcal{R}, \sigma) = y \int_y^\infty \frac{{\cal N}(x | \mathcal{R}, \sigma)}{x (x^2 - y^2)^{1/2} } dx
,\end{equation}
where $x \equiv R$, and ${\mathcal{N}}(x | \mathcal{R}, \sigma)$ is the normal distribution with expected value $\mathcal{R}$ and standard deviation $\sigma$; 
the first two moments of the distributions are related by
\begin{eqnarray}
\label{eq:moments}
\bar{x} &=& \frac{4}{\pi} \bar{y}; \\ \nonumber
\bar{x^2} &=& \frac{3}{2} \bar{y^2}. \\ \nonumber
\end{eqnarray}

As there is no known analytical solution to the integral in Eq.\,\eqref{eq:pdf}, this is evaluated numerically.
From Eq.\,\eqref{eq:pdf} the cumulative distribution function (c.d.f.)
\begin{equation}
\label{eq:cdf}
\Phi(y|\mathcal{R},\sigma) = \int_0^y \phi(y'|\mathcal{R}, \sigma) dy'
\end{equation}
is also estimated numerically.
Hereafter the dependence on the parameters $\mathcal{R}$ and $\sigma$ is considered implicitly, that is, $\phi(y) = \phi(y| \mathcal{R}, \sigma)$ and $\Phi(y) = \Phi(y| \mathcal{R}, \sigma)$.

\subsection{Censoring and truncation}

The finite wavelength resolution of spectrographs and the intrinsic broadening of a non-rotating star set limits on the capability of measuring \vsini\ below a certain threshold. 
This can be due to low $v_{\rm eq}$, low $i$, or both.
When $i$ is sufficiently low, we can also expect that $P$ cannot be measured because the star is seen almost pole-on and the rotational modulation induced by surface inhomogeneities is therefore undetectable. 
In practice, $P$ is still measurable also when only a \vsini\ upper limit can be estimated, and therefore the limits on \vsini\ in general dominate those on $P$. 
Other cases when $P$ is expected not to be measurable include uniformly distributed surface inhomogeneities and unfavourable photometric sampling.

A survival analysis \citep[][]{Klein+2003,2012msma.book.....F} can be applied to recover $\mathcal{R}$ and $\sigma$ from a $\{R \sin i\}$ set in presence of censoring and truncation, in which cases Eq.\,\eqref{eq:moments} become invalid.
When only a \vsini\ upper limit, $(v \sin i)_{\rm lim}$, is available for the object, this is translated into an $R \sin i$ upper limit using the relation
\begin{equation}
\label{eq:RsiniLim}
(R \sin i)_{\rm lim} = \mathcal{C} P (v \sin i)_{\rm lim}
,\end{equation}
which is then considered as a left-censored data point.

At the other extreme, ultra-fast rotators may have such a high $v \sin i$ value that its measurement is very uncertain or impossible, and therefore only a lower limit is available.
These measurements can be treated as right-censored data points, for which a $(R \sin i)$ lower limit can be defined in analogy with Eq.\,\eqref{eq:RsiniLim}. 

Left- and right-censored data points are taken into account through a survival function \citep{Klein+2003}, which gives the probability that an object has a value above some specified level.
For the case at hand, the survival function is
\begin{equation}
\label{eq:survival}
S(y) = P( R \sin i > y) = 1 - \Phi(y),
\end{equation}
where $\Phi(y)$ is given by Eq.\,\eqref{eq:cdf}.
Using Eq.\,\eqref{eq:survival}, we assign a probability $S(y)$ if $y$ is a right-censored (lower limit) data point and $(1-S(y))$ if $y$ is a left-censored (upper limit) data point.

The incompleteness of the sample does not represent a limitation for the analysis as long as selection effects do not depend on the variates of interest.
We expect, however, that when $i$ is sufficiently low, neither \vsini\ nor $P$ can be measured and that the data are left-truncated in most cases.
Data points like this are not present in the $\{R \sin i\}$ dataset, but we do not know the exact truncation value $(R \sin i)_{\rm trunc}$ in each bin a priori. 
However, for a random orientation of the spin axis the probability of occurrence of the inclination $i$ between $i$ and $i + di$ is $\sin i\, di,$ and therefore the distribution favours high values of $i$.
The results are therefore quite insensitive to the accuracy in determining the left-truncation level, and we adopt the practical approach of considering the lowest $R \sin i$ value as an approximation of $(R \sin i)_{\rm trunc}$ in each bin.
The way in which truncation is taken into account in the analysis is described in Sect.\,\ref{eq:MeanRadius}. 

Right-truncation is ignored because we assumed that for ultra-fast rotator for which the \vsini\ measurement is uncertain or impossible a lower limit is reported instead of omitting the measurement.

\subsection{Estimating the mean radius from the projected radius distribution}
\label{eq:MeanRadius}

After the p.d.f. and the corresponding c.d.f. and survival function (Eqs.\,\eqref{eq:pdf}, \eqref{eq:cdf}, and \eqref{eq:survival}) were defined, we defined a likelihood function as
\begin{equation}
\label{eq:likelihood}
\mathcal{L} = \prod_{\rm det}\phi(y_j) \prod_{\rm lcens}\left[1-S(y_j)\right] \prod_{\rm rcens} S(y_j)
,\end{equation}
where the products are over detected, left-censored, and right-censored data points, respectively.
Since our dataset is always left-truncated, in Eq.\,\eqref{eq:likelihood} we replace $\phi(y_j)$ by 
$\phi(y_j)/S(Y_L)$
and $S(y_j)$ by
$S(y_j)/S(Y_L)$ \citep{Klein+2003},
with $Y_L = (R \sin i)_{\rm trunc}$ the left-truncation $y$ value in the dataset. 
The negative log-likelihood function is therefore
\begin{equation}
\begin{split}
\label{eq:minusloglikelihood}
- & \ln \mathcal{L} = \\ 
- & \left( \sum_{\rm det} \ln(\phi(y_j)) + \sum_{\rm lcens} \ln(1-S(y_j)) + \sum_{\rm rcens} \ln(S(y_j)) \right).
\end{split}
\end{equation}
The parameters $\mathcal{R}$ and $\sigma$ are finally evaluated by minimising the negative log-likelihood function Eq.\,\eqref{eq:minusloglikelihood} using the \texttt{L-BFGS-B} optimisation method of \cite{Byrd_etal:95} as implemented in \texttt{R}.

\subsection{Simulations}
\label{sec:simulations}

The method was extensively tested using numerical simulations.
For brevity, only a summary of these tests is reported here.
More details are reported in Appendix\,\ref{sec:SimuResults}.

For a sufficiently low level of censoring and truncation, that
is, $(R \sin i)_{\rm trunc} \apprle 0.1 \mathcal{R}$ and $(R \sin i)_{\rm cens} \apprle 0.3 \mathcal{R}$, and $\sigma / \mathcal{R} \approx 0.1$ the method is capable of reproducing $\mathcal{R}$ within $\approx$ 2 percent and $\sigma$ within $\approx$ 2 percent with $n \apprge 10$ data points (median over 100 realisation - see Appendix\,\ref{sec:SimuResults}).

Maintaining a low level of censoring and truncation, the accuracy degrades for extremely high or low values of $\sigma / \mathcal{R}$. 
For $\sigma \rightarrow 0$, the numerator in Eq.\,\eqref{eq:pdf} tends to a Dirac delta function and $\phi$ develops a singularity at $y = \mathcal{R}$ \cite[see][]{1950ApJ...111..142C}.
As a consequence, for $\sigma/\mathcal{R} \apprle 0.01$ the numerator of the integrand in Eq.\,\eqref{eq:pdf} is a very steep function of $x$, which causes numerical instabilities in the evaluation of the integral.
Simulations for $\sigma/\mathcal{R} \rightarrow 0.01$ show that in this limit $\mathcal{R}$ can still be recovered within $\approx$ 5 percent, while $\sigma$ can be overestimated by a factor of several because of the smoothing implied by the integral in Eq.\,\eqref{eq:pdf}.
At the other extreme, for $\sigma/\mathcal{R} \apprge 0.3$ the theoretical distribution would imply a non-negligible probability of having negative (unphysical) values and the numerical procedure fails.
Simulations for $\sigma/\mathcal{R} \rightarrow 0.3$ show that in such extreme cases $\mathcal{R}$ can be recovered within $\approx$ 4 percent and $\sigma$ within $\approx$ 50 percent (median over 100 realisations).

Fixing $\sigma / \mathcal{R} \approx 0.1$, an increase in the level of censoring and truncation does not significantly decrease the accuracy in reproducing $\mathcal{R}$ and $\sigma$ as long as the core of the distribution remains unaffected.
In practice, both $\mathcal{R}$ and $\sigma$ are recovered within a few percent up to $(R \sin i)_{\rm trunc}$ and $(R \sin i)_{\rm cens} \approx 0.6 \mathcal{R}$.
Above this level, censoring and truncation affect the core of the distribution, making it difficult to recover its original shape.

We note that these simulations outline limitations that are intrinsic to the problem at hand and cannot be overcome using a simplified approach, and they set the boundaries for a meaningful estimate of the mean radius from an $\{R \sin i\}$ set.

\section{Mean stellar radii in the Pleiades}
\label{sec:Pleiades}

\begin{figure*}[ht]
\begin{center}
\includegraphics[width=0.4\textwidth]{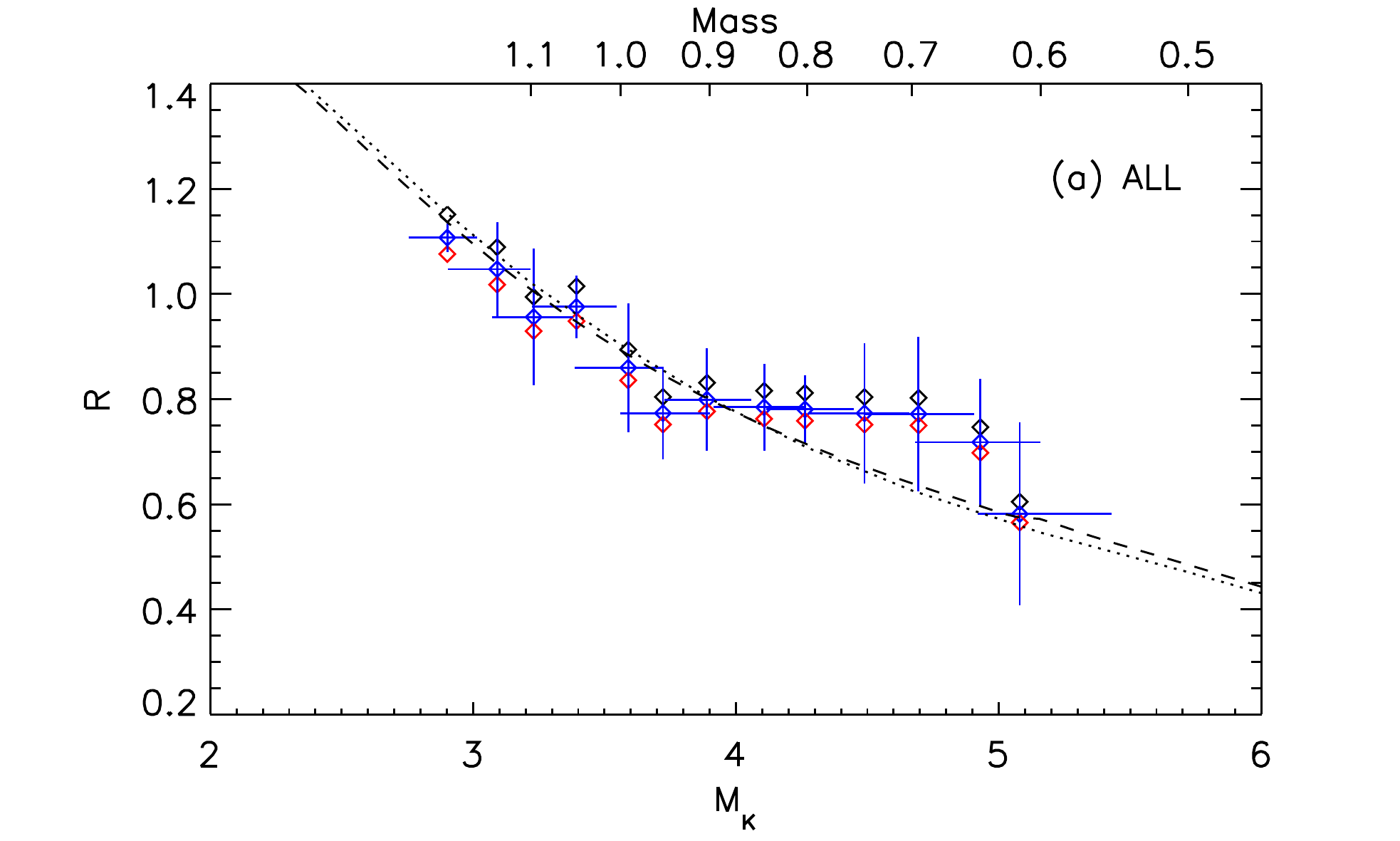}
\includegraphics[width=0.4\textwidth]{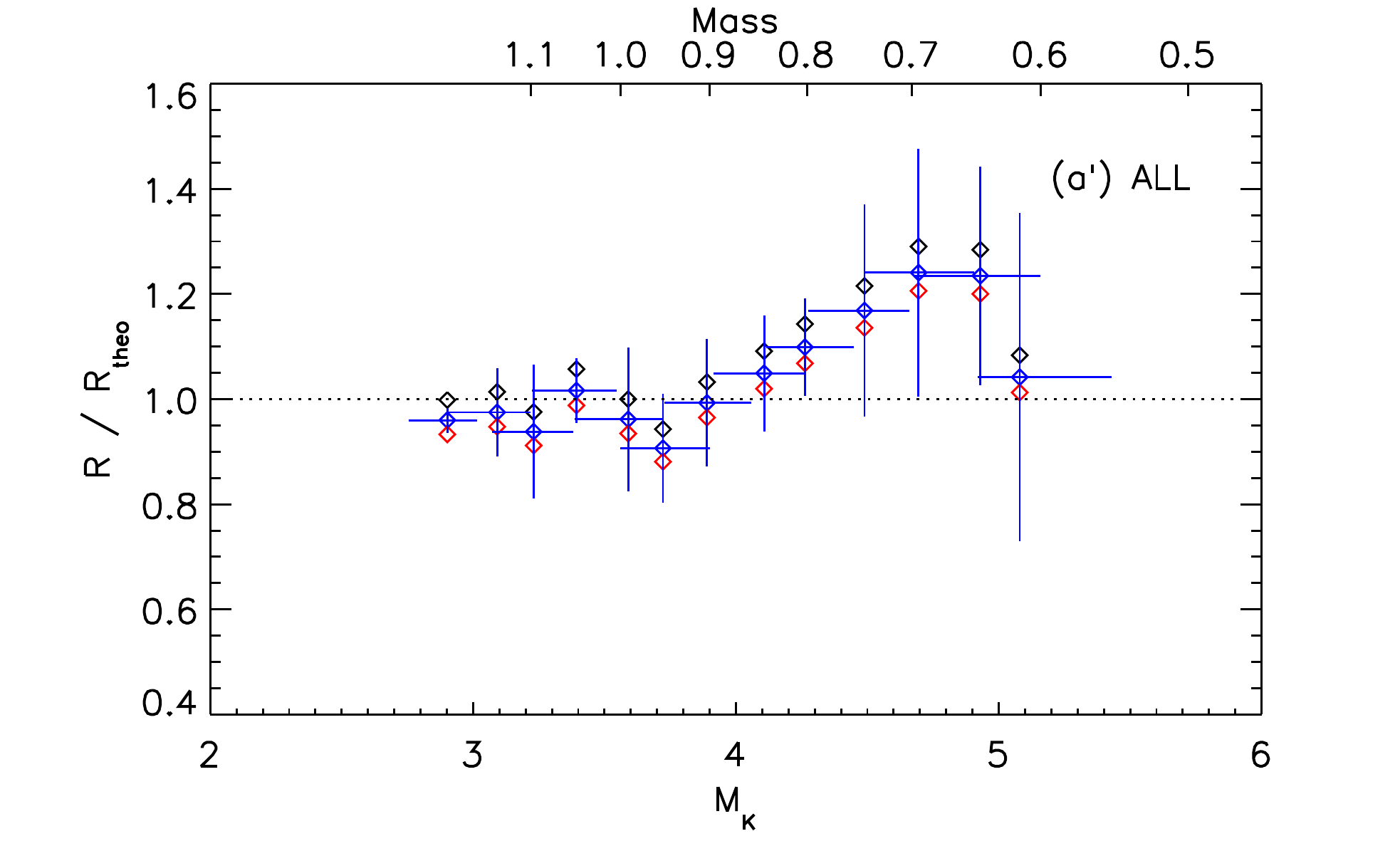}
\includegraphics[width=0.4\textwidth]{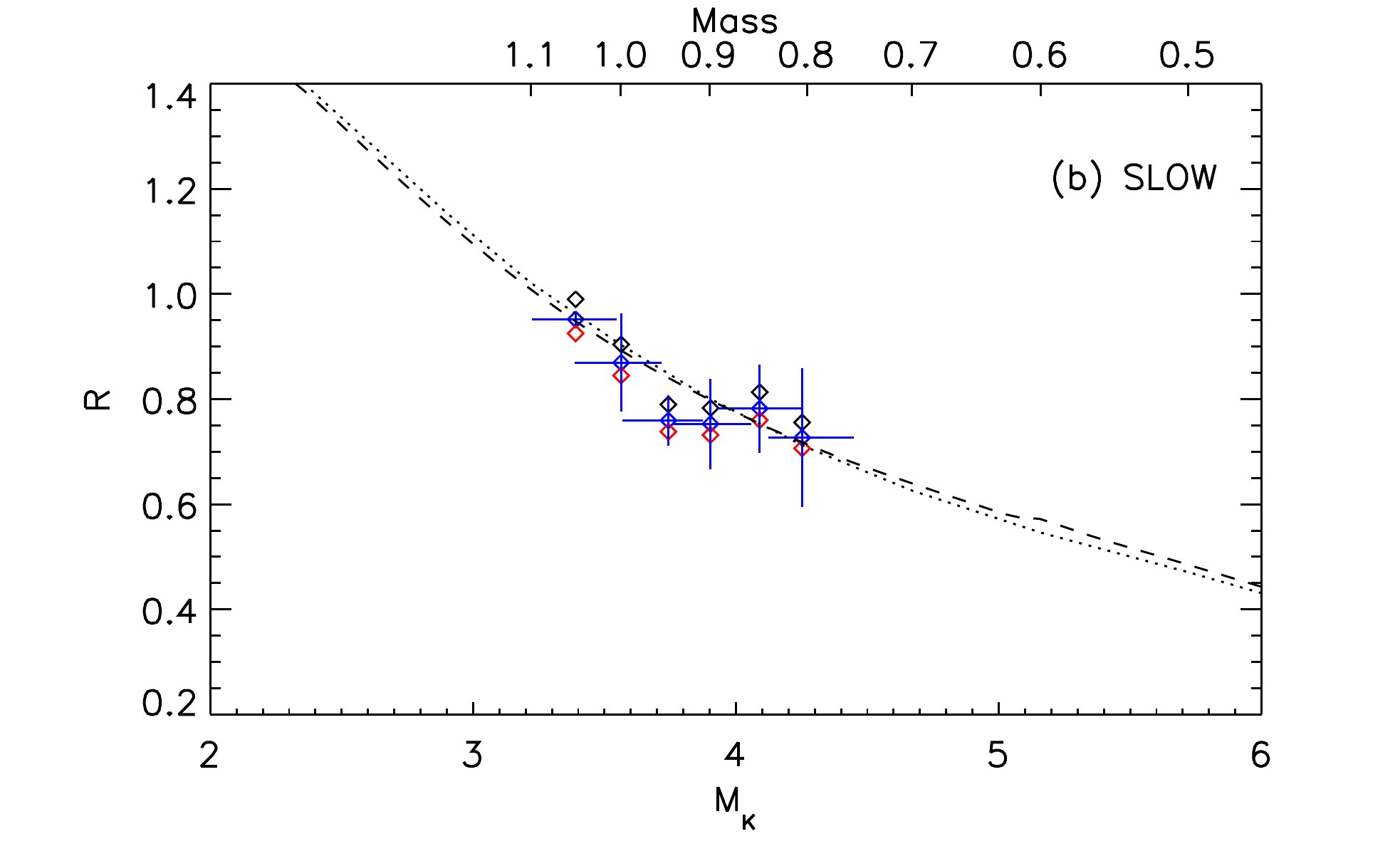}
\includegraphics[width=0.4\textwidth]{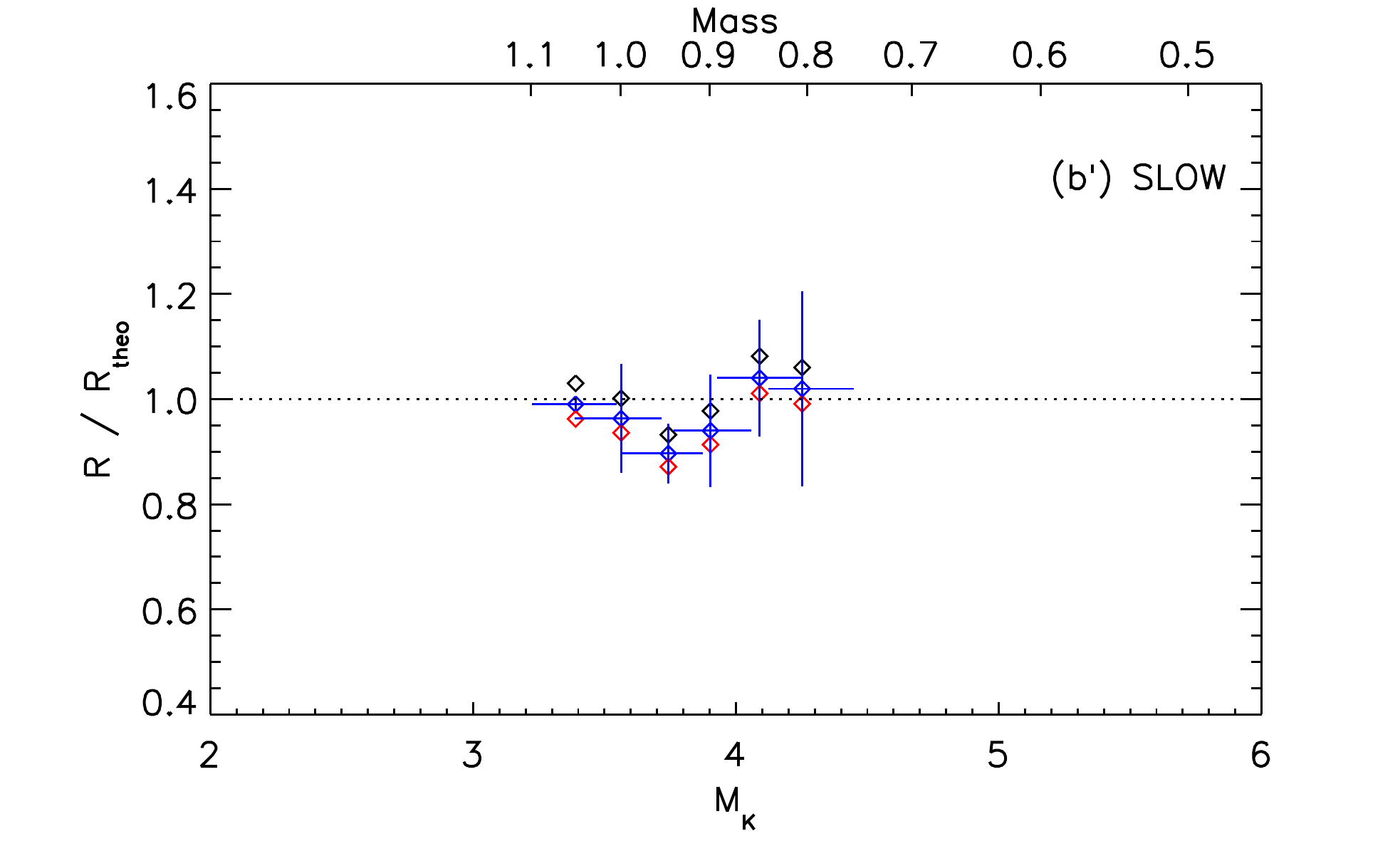}
\includegraphics[width=0.4\textwidth]{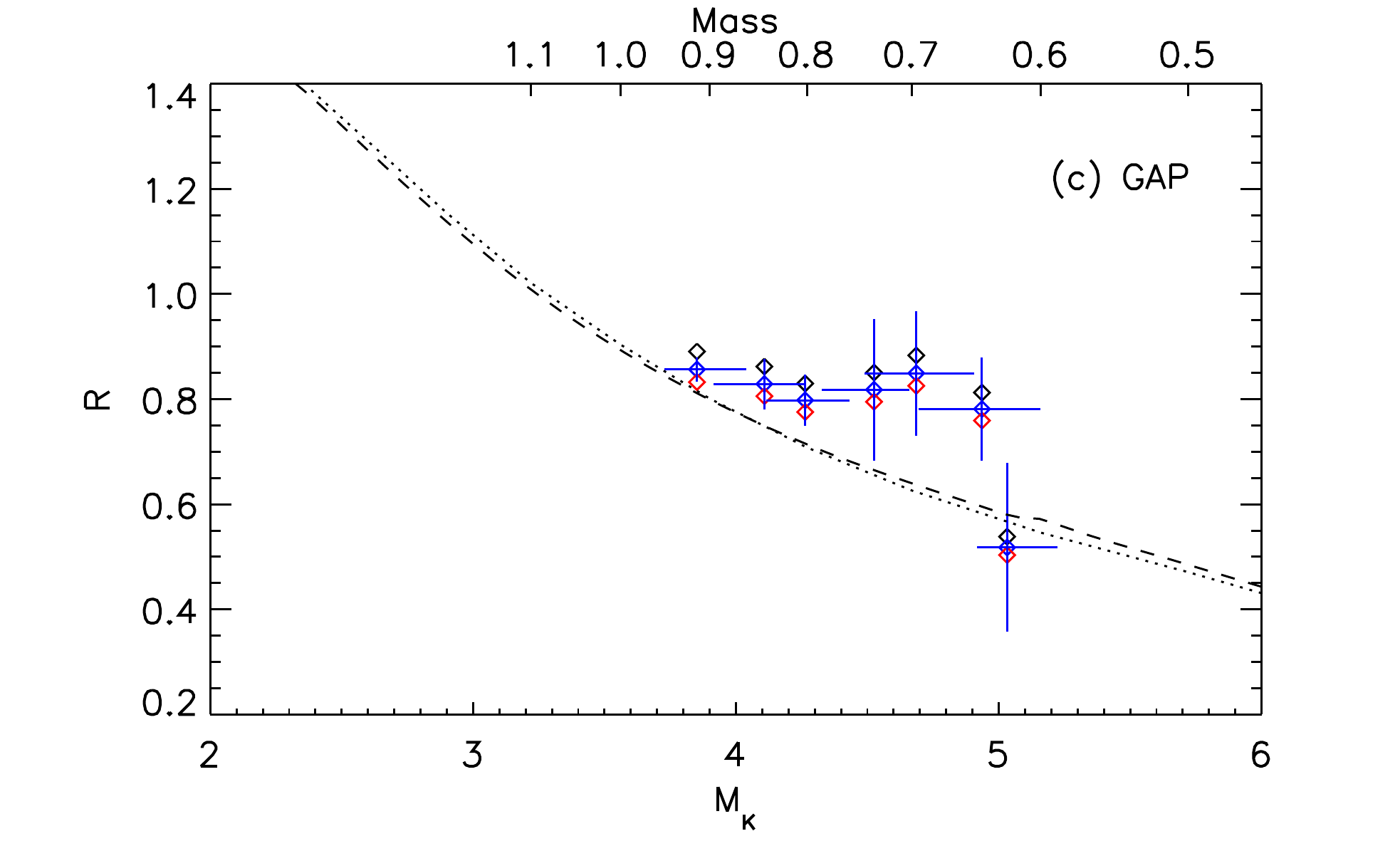}
\includegraphics[width=0.4\textwidth]{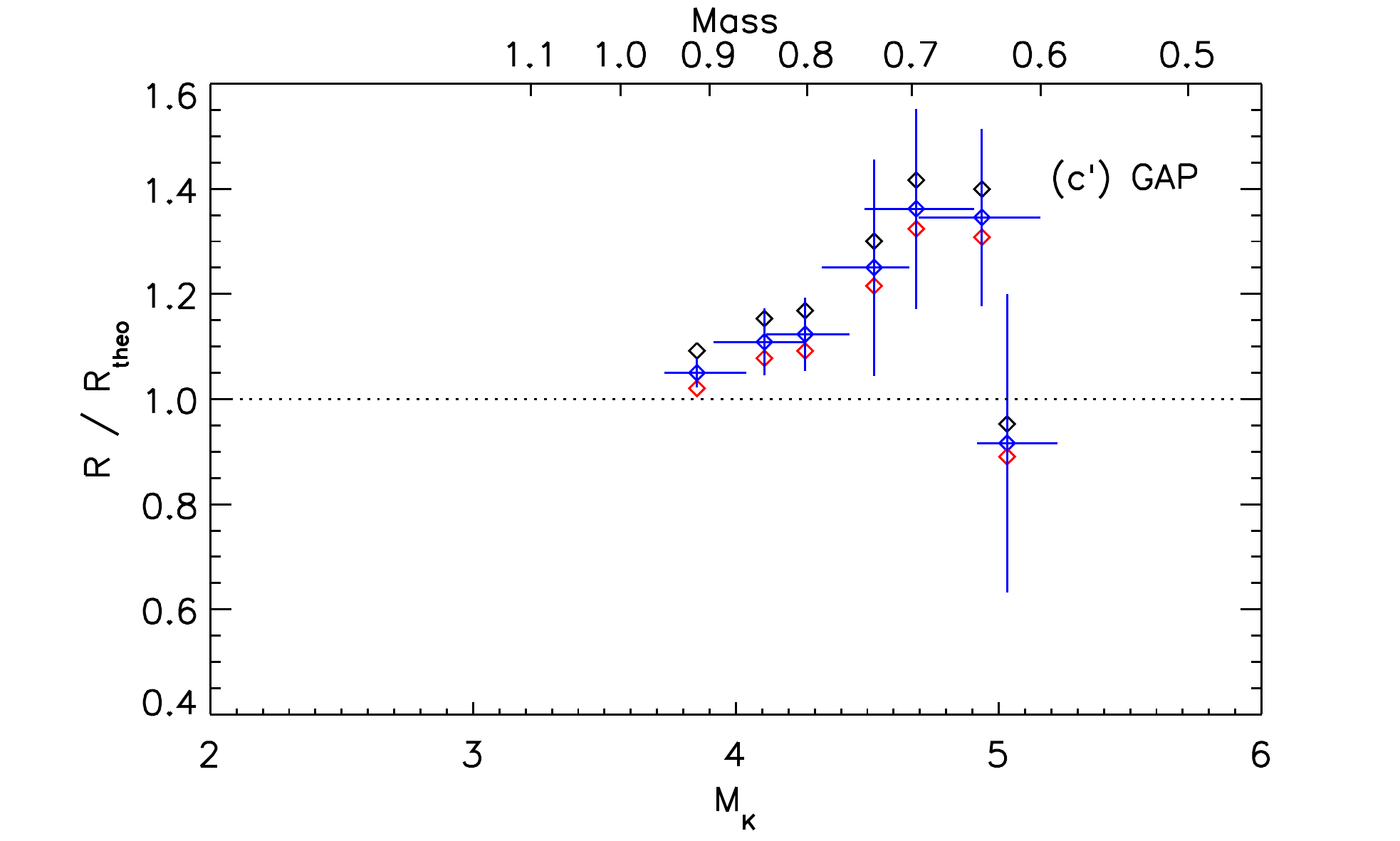}
\includegraphics[width=0.4\textwidth]{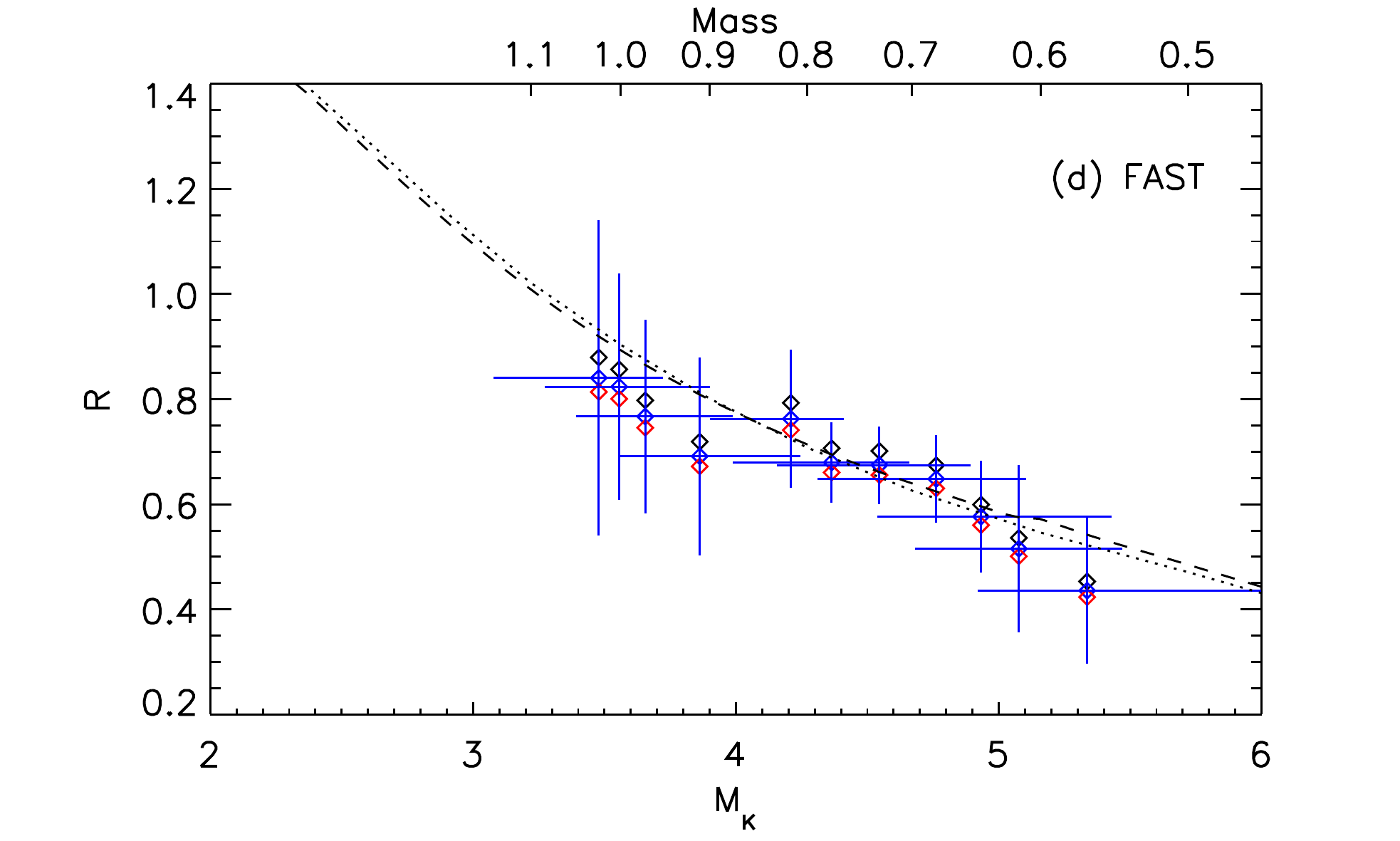}
\includegraphics[width=0.4\textwidth]{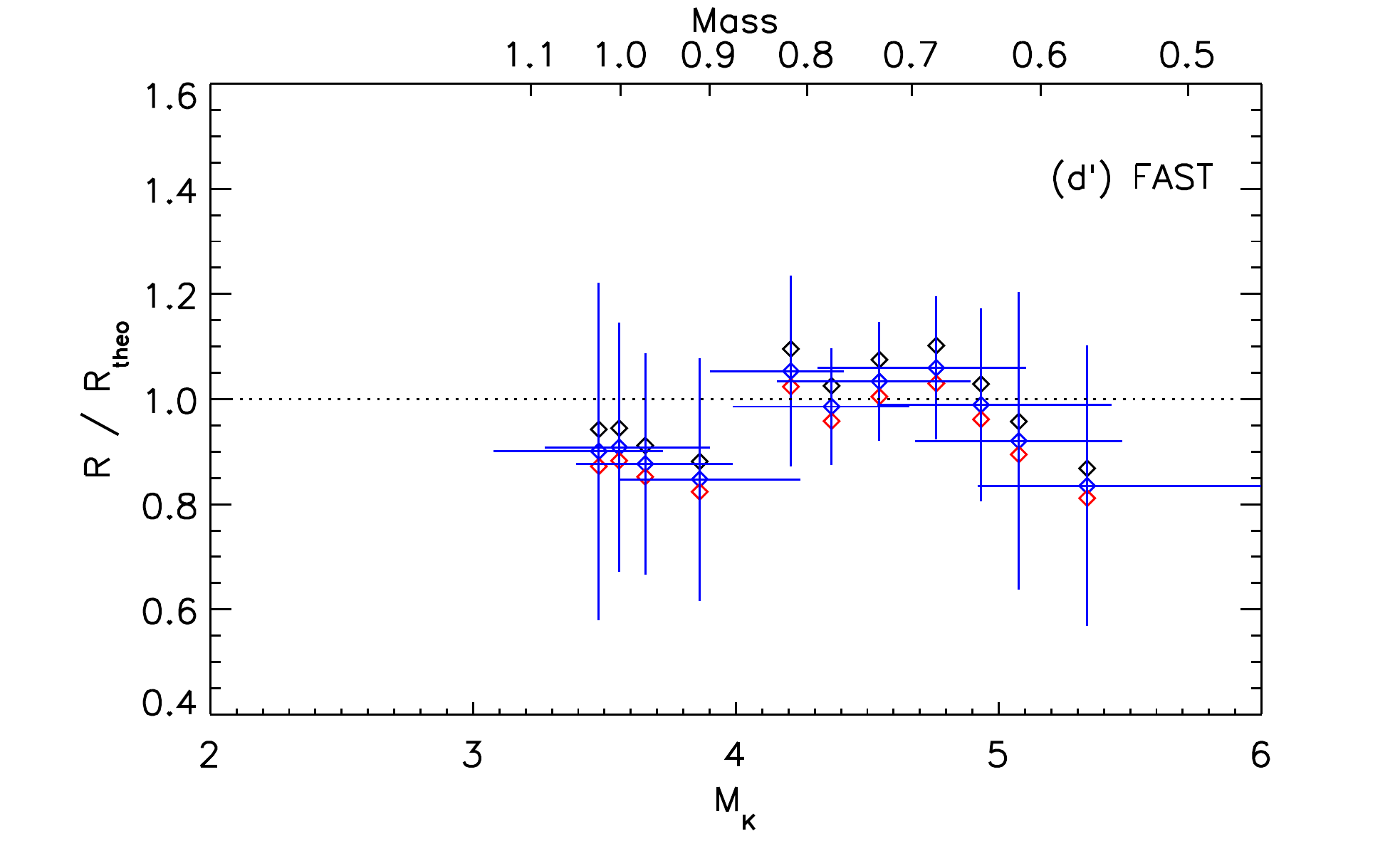}
\caption{Comparison of the expected radius $\mathcal{R}$ with the models of \cite{2015A&A...577A..42B} (dotted line) and \cite{2013ApJ...776...87S} (dashed line).
$\mathcal{R}$ is reported as a function of the average $M_K$ in each bin.
The three values of $\mathcal{R}$ correspond to our best estimate of the SDR (blue diamond) and the estimated upper and lower SDR limits (black and red diamonds).
The horizontal bars encompass the $M_K$ range in each bin.
The vertical bars represent the standard deviation $\sigma$ for each bin centred on our best estimate of the SDR.
Right panels show the ratio with the \cite{2015A&A...577A..42B} model.
Panels (a) and (a') are obtained using all data, panels (b) and (b') for the slow-rotator sequence, panels (c) and (c') for stars with periods between fast and slow rotators, panels (d) and (d') for fast rotators only ($P < 2$ d). Bins are $\Delta M \approx 0.1 M_{\odot}$ wide except for the fast rotators set, for which $\Delta M \approx 0.2 M_{\odot}$ to ensure that at least $\approx$10 stars fall in each bin.
}
\label{fig:Pleiades_rout}
\end{center}
\end{figure*}

The method described in Sect.\,\ref{sec:Method} was applied for the first time to the dataset available for the Pleiades (Sect.\,\ref{sec:data}).
The data were organised in $M_K$ bins corresponding to fixed $\Delta M$ according to the theoretical relationship reported
by \cite{2015A&A...577A..42B} for $M_K$ vs $M$  at the Pleiades age.
For each bin we compute $\mathcal{R}$ and $\sigma$ from the $R \sin i$ distribution. 
In Fig.\,\ref{fig:Pleiades_rout} we compare $\mathcal{R}$ vs $\left<M_K\right>$ with the models of \cite{2015A&A...577A..42B} and \cite{2013ApJ...776...87S}.
The comparison is carried out for the whole $P$ range, for the slow-rotator sequence \citep[as identified in ][]{2015A&A...584A..30L}, for the fast rotators ($P < 2$~d), and for the stars with periods in between the fast rotators and the slow-rotator sequence.
According to the definition of \cite{Barnes:2003}, these correspond to the I sequence, the C sequence and stars in the gap\footnote{We
note that 14 stars of the whole sample are not assigned to any sequence as their period is at least 1$\sigma$ above the slow-rotator sequence.}.
Bins are $\Delta M \approx 0.1 M_{\odot}$ wide, except for fast rotators, for which $\Delta M \approx 0.2 M_{\odot}$ to have at least ten stars in each bin, and which are allowed to overlap to provide $\mathcal{R}$ estimates at steps of $\approx 0.05 M_{\odot}$.
For $1.2 \apprge M/M_{\odot} \apprge 0.85 $, using the whole $P$ range, $\mathcal{R}$ agrees with the theoretical models within $\approx$10 percent ($\approx$5 percent for $1.2 \apprge M/M_{\odot} \apprge 1.00 $) with no significant bias.
The noise in the data, however, increases with decreasing mass, resulting in increasingly larger $\sigma$.
For $0.85 \apprge M/M_{\odot} \apprge 0.65 $, $\mathcal{R}$ becomes systematically larger than theory, although still within $\approx 1 \sigma$.
Restricting the calculations to stars belonging to the slow-rotator sequence, $\mathcal{R}$ still agrees with the theoretical models within $\approx$10 percent with no significant bias.
The results for the fast rotators have larger $\sigma$ and $\mathcal{R}$ scatter, up to $\approx$ 15 percent, mostly because larger bins are required to have at least ten data points in each bin, but no systematic deviations from the models are found.
The discrepancy with the theoretical model is confined to stars with mass just below the low-mass end of the slow-rotator sequence and $P > 2$~d (panels (c) and (c') in Fig.\,\ref{fig:Pleiades_rout}), for which $\mathcal{R}$ is inflated at 2$\sigma$ level.
According to the scenario described by \cite{Barnes:2003}, these are stars that converge on the slow-rotator sequence (gap).

Possible alternative explanations for this behaviour are
\begin{enumerate}
\item{\label{one}systematic deviations due to some outliers;}
\item{\label{three}$P$-dependent observational biases not taken into account by our procedure;}
\item{\label{four}biases in the \vsini\ datasets.}
\end{enumerate}
To consider the first alternative explanation, we repeated the calculation using different binning and obtained essentially the same results.
We also repeated the calculations excluding $m = 3$ stars in each bin and using all $n-m$ combinations, with $n$ the number of stars in each bin, with no significant change with respect to the original results.
In this way we verified that there is indeed a group of measurements that produces the observed deviation and not just some isolated outliers, to which our procedure is rather insensitive in any case.

The second alternative explanation is deemed rather unlikely since the decrease in fraction of stars with both $P$ and \vsini\ measurements is rather uniform in $P$ down to at least $M \approx 0.6 M_{\odot}$ (see Fig.\,\ref{fig:data}) and the $P$ dataset of \cite{Hartman_etal:2010}  is 93 percent complete in the mass range $1.0 \apprge M/M_{\odot} \apprge 0.7$.  

The third alternative explanation is therefore the only one that could be of some concern.
\cite{Hartman_etal:2010} estimated that for $M \apprle 0.85 M_{\odot}$ the \vsini\ dataset could be affected by a bias $\Delta v \sin i \approx -1.5$ km s$^{-1}$, but this was still insufficient to explain the $\sin i$ distribution, so that the authors invoked other factors like the radius inflation and a rather high SDR. 
By constraining the SDR effects as described in Sect.\,\ref{sec:SDR}, we estimate that to explain the $\mathcal{R}$ vs $\left<M_K\right>$ discrepancy at $M\approx 0.7 M_{\odot}$, a bias of at least $\Delta v \sin i \approx -4.0$ km s$^{-1}$ limited to a rather restricted $M_K$ range would be necessary.
Considering the expected uncertainties in the intrinsic width of non-rotating stars \citep[e.g.][]{1998A&A...335..183Q}, which are expected to be the main cause of such systematic deviations, this bias seems too high and its dependence on spectral type much steeper than deemed plausible.

\begin{table}
   \centering
      \caption{Ratio between the expected radius $\mathcal{R}$ and the theoretical average radius $\left<R\right>$ in a $\Delta M \approx 0.2 M_{\odot}$ bin centred on $M \approx 0.7 M_{\odot}$ for stars not belonging to the slow-rotator sequence and $P>2$~d. Calculations are carried out using all \vsini\ values, only the \cite{1998A&A...335..183Q} \vsini\ (Q98), and only the \cite{2000AJ....119.1303T} \vsini\ (T00). $n$ is the number of points in the bin, $n_{\rm ul}$ the number of \vsini\ upper limits. The estimated uncertainties due to SDR are reported together with $\mathcal{R} / \left<R_{\rm theo}\right>$, with a separate column for the standard deviation $\sigma$ scaled by $\left<R_{\rm theo}\right>$. }
      \label{tab:inflation}
      \begin{tabular}{rrrrrr}
      \hline
Set & $\left<M\right>$ & $n$ & $n_{\rm ul}$ & $\bar{R} / \left<R_{\rm theo}\right>$ & $\sigma / \left<R_{\rm theo}\right>$ \\
\hline
All & 0.68 & 41 & 5 & $1.31_{-0.04}^{+0.05}$ & 0.19 \\
Q98 & 0.71 & 24 & 0 & $1.33_{-0.04}^{+0.05}$ & 0.17 \\
T00 & 0.65 & 13 & 3 & $1.14_{-0.03}^{+0.04}$ & 0.12 \\
\hline
      \end{tabular}
\end{table}

As discussed in Sect.\,\ref{sec:data}, \vsini\ of fast rotators are mostly adopted from \cite{1993ApJS...85..315S}, while for the other stars they are mostly taken from \cite{1998A&A...335..183Q} for $M \apprge 0.85 M_{\odot}$ and from \cite{2000AJ....119.1303T} for $M < 0.85 M_{\odot}$.
Excluding the fast-rotator measurements of \cite{1993ApJS...85..315S}, the comparison for the five stars in common between the two remaining datasets does not point to any significant bias. 
To investigate the possible \vsini\ bias in more detail,
we repeated the calculations on a $\Delta M \approx 0.2 M_{\odot}$ bin centred on $M \approx 0.7 M_{\odot}$ using all available \vsini, only the Q98, and only the T00 \vsini.
The results, reported in Table\,\ref{tab:inflation}, show that in all three cases the expected mean radius is larger than the theoretical mean radius by 1 or 2 $\sigma$.
We note that the discrepancies are smaller than those shown in Fig.\,\ref{fig:Pleiades_rout} (panels (c) and (c')) because to have at least ten stars for each set, the bin is larger and the mean mass in two sets (Q98 and T00) differs by 0.06 $M_{\odot}$.
In conclusion, we have no evidence of an observational \vsini\ bias at the level required to explain the radius discrepancy found for stars that are converging on the slow-rotator sequence. 

From a different perspective, it can
be argued that the results obtained for the fast-rotator sequence may be affected by the possible omission of high \vsini\ values that have not been reported as lower limits as our method requires.
This aspect can be of concern as more than one hundred stars of the \cite{Hartman_etal:2010} periods sample have no \vsini\ measurement and a significant fraction of them have short periods.
In the Pleiades case, however, we can reasonably assume that the lacking \vsini\ data do not depend on its value as high values of \vsini\ and lower limit are reported.
The treatment of datasets in which high \vsini\ values are lacking would require sufficient information to allow considering right-truncation in Eq.\,\eqref{eq:likelihood}.

\section{Conclusions}
\label{sec:Conclusions}

We have set up a new method for deriving mean stellar radii from rotational periods, $P$, and projected rotational velocities, \vsini, based on the survival analysis concept \citep{Klein+2003}.
This method exploits the whole information content of the dataset with an appropriate statistical treatment of censored and truncated data. 
Provided censoring and truncation do not significantly affect the peak of the $R \sin i$ distribution and that there is no significant bias in the data, the method can recover the mean stellar radius with an accuracy of a few percent with as few as $n \approx 10$ measurements.
The total standard deviation, $\sigma$, which cumulatively takes the data noise and the intrinsic $R$ standard deviation into account, can also be estimated with an accuracy of a few percent except in extreme cases where the distribution is too broad ($\sigma / \mathcal{R} \sim 0.3$) or too narrow ($\sigma / \mathcal{R} \sim 0.01$).

The method has been applied for the first time to the dataset available for the Pleiades.
We found that deviations of the empirical $\mathcal{R}$ vs $M_K$ relationship from standard models \citep[e.g.][]{2013ApJ...776...87S,2015A&A...577A..42B} do not exceed 5 percent for $1.2 \ge M/M_{\odot} \ge 1.0$ and 10 percent for $1.0 > M/M_{\odot} \ge 0.85$, with no significant bias.
Evidence of a systematic deviation at $1-2\sigma$ level of the empirical $\mathcal{R}$ vs $M_K$ relationship from standard models is found only for stars with $M \approx 0.7 \pm 0.1 M_{\odot}$ that are converging on the slow-rotator sequence.
Deviations of the $\mathcal{R}$ vs $M_K$ relationship for fast rotators ($P<2$~d) do not exceed $\approx 15$ percent in the whole mass range with no evidence of a systematic deviation from standard models.
No evidence of a radius inflation of fast rotators in the Pleiades is therefore found.


\begin{acknowledgements}

FS acknowledges support from the Leibniz Institute for Astrophysics Potsdam (AIP) through the Karl Schwarzschild Postdoctoral Fellowship.
Research at the Universit\`a di Catania and at INAF - Osservatorio Astrofisico di Catania is funded by MIUR (Italian Ministry of University and Research).
The authors warmly thank Sydney Barnes (Leibniz Institute for Astrophysics Potsdam) for valuable discussions and an anonymous referee for useful comments.
This research made use of the NASA Astrophysics Data System Bibliographic Services (ADS) and the \texttt{R} software environment and packages (\url{https://www.r-project.org}).
\end{acknowledgements}

\bibliographystyle{aa}
\bibliography{RsiniMethod}

\appendix
\section{Accuracy of the method}
\label{sec:SimuResults}

In this appendix we present some examples and tests carried out to estimate the method accuracy.

\begin{figure*}[ht]
\begin{center}
\includegraphics[width=0.44\textwidth]{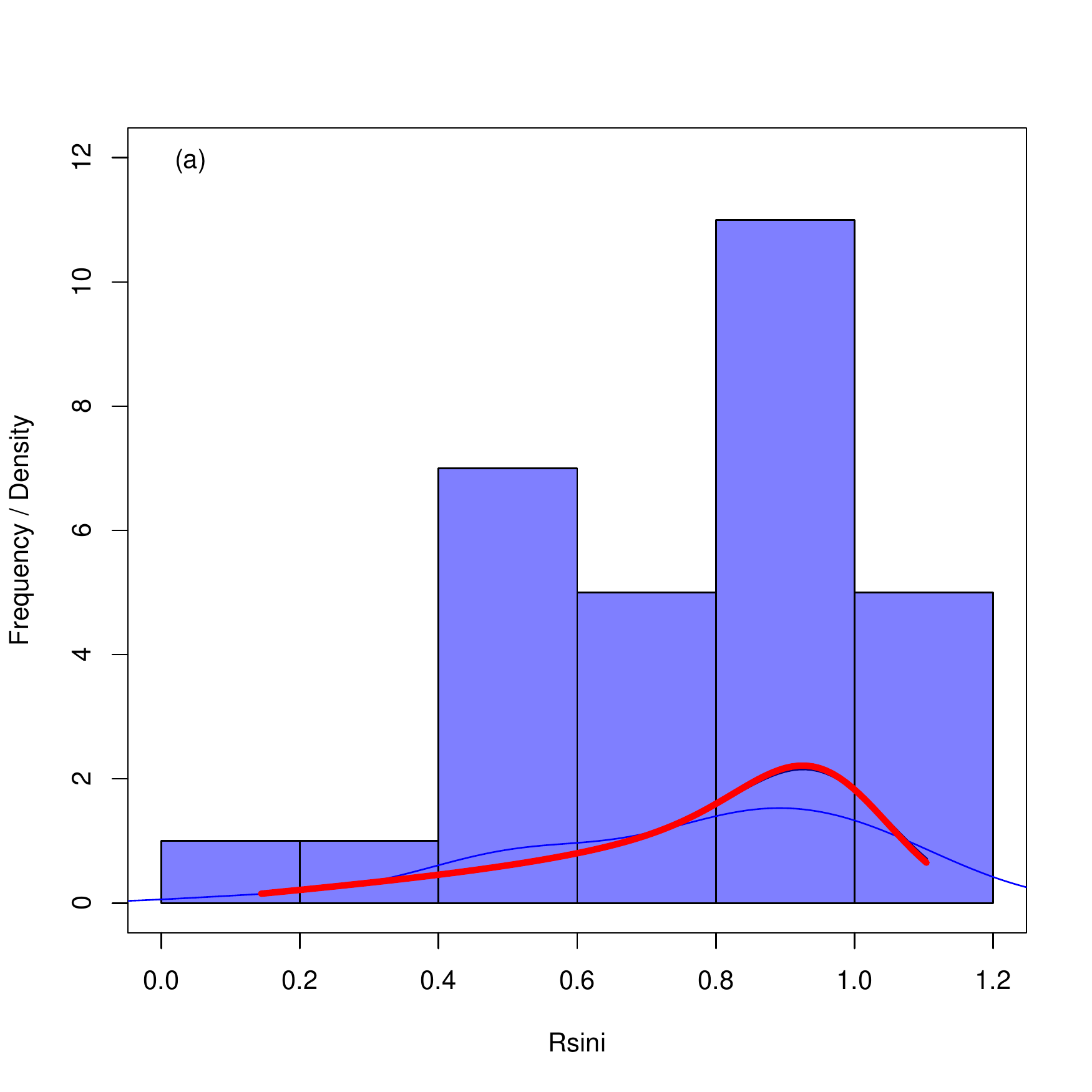}
\includegraphics[width=0.44\textwidth]{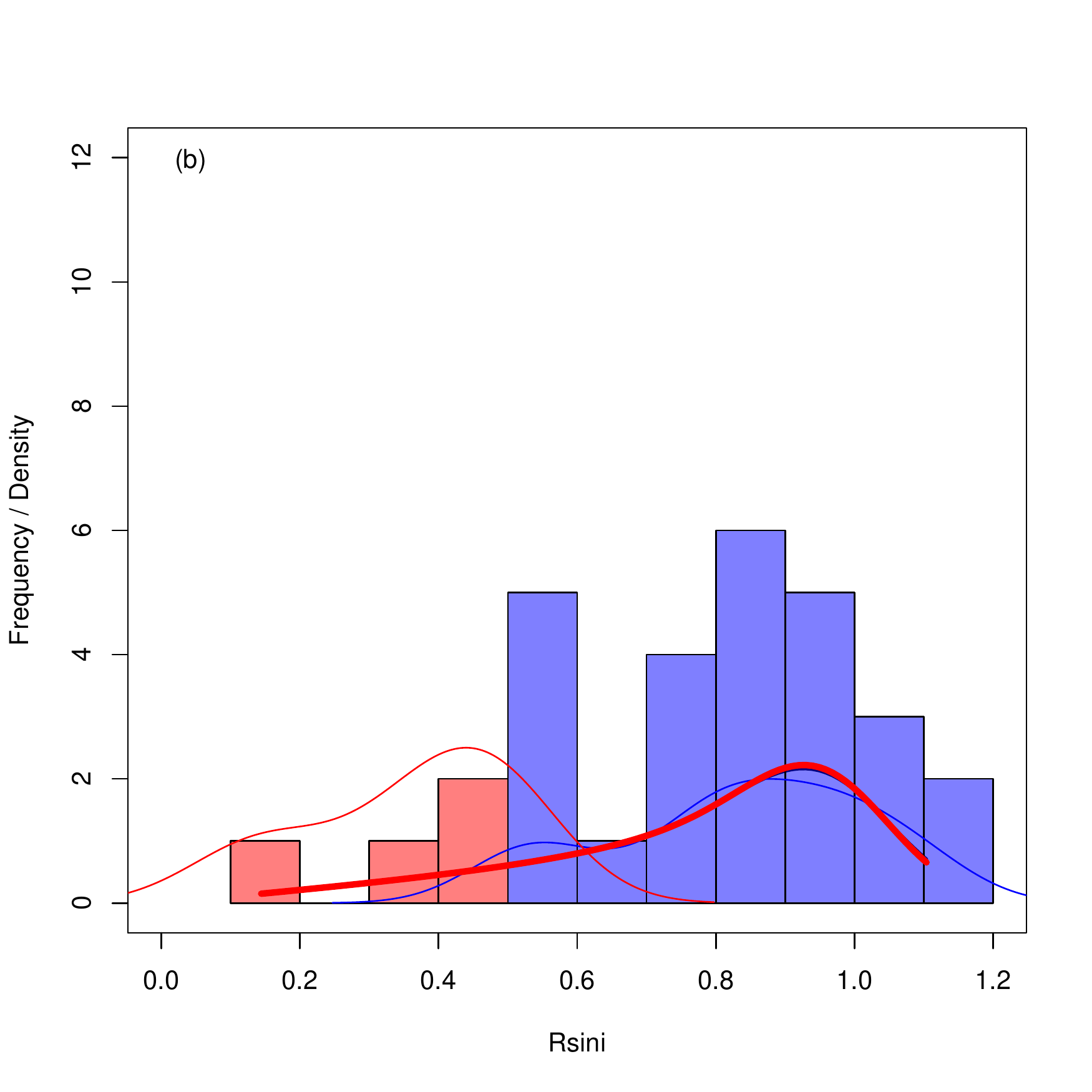}
\includegraphics[width=0.44\textwidth]{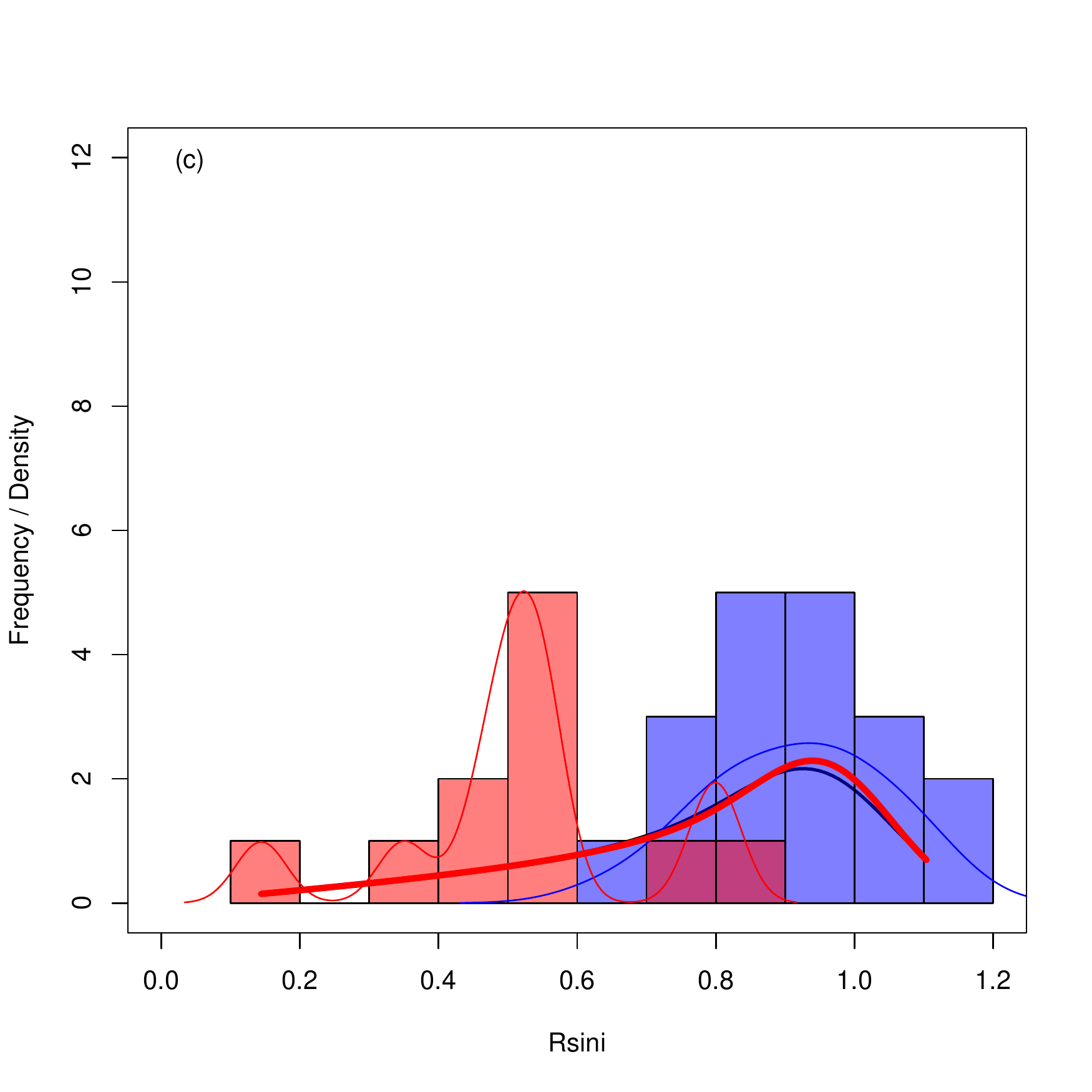}
\includegraphics[width=0.44\textwidth]{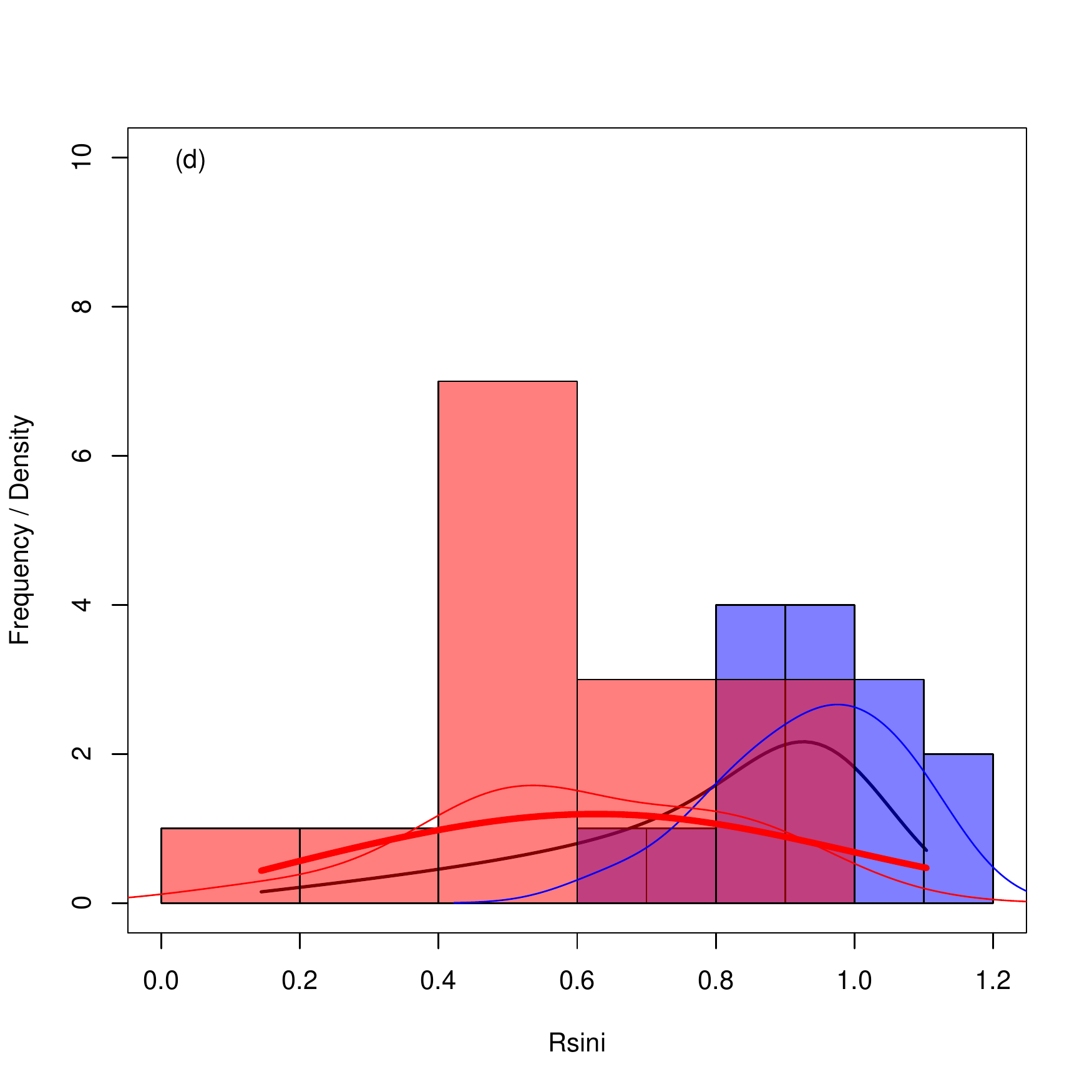}
\caption{Illustrative tests in recovering $\mathcal{R}$ from a synthetic $\{ R \sin i\}$ set with censoring.
The simulations reproduce $n=30$ slow-rotator sequence stars  in the Pleiades with $M\sim 1M_{\odot}$ and $R\sim \ 1 R_{\odot}$ at varying censoring levels: $(v \sin i)_{\rm lim} = 0$ (panel a);  6 (panel b); 9 (panel c); 10 (panel d) km s$^{-1}$.
Blue and red histograms are used for censored and not-censored data.
The kernel density estimate is shown as a thin blue line for not-censored data and as a red thin line for censored data, the function $\phi$ corresponding to the given synthetic distribution parameters is plotted as a thick black line and that corresponding to the reconstructed parameter as a thick red line.
Histograms are representations of frequency.
}
\label{fig:test_synth_censored}
\end{center}
\end{figure*}

We considered as an illustrative example a group of stars with $M\sim 1 M_{\odot}$ in the Pleiades slow-rotator sequence \citep{2015A&A...584A..30L}. 
This group of stars has a normal distribution of periods with $\bar{P} \simeq 3.53$ d and $\sigma_P \simeq 0.35$ \cite[see Table 2 in ][]{2015A&A...584A..30L}. 
Assuming $R \sim 1 R_{\odot}$, it follows that the equatorial velocities have a normal distribution with  $\bar{v_{\rm eq}} \simeq 12.7$ and $\sigma_v \simeq 1.3$ km\,s$^{-1}$.
Assuming that this group is composed of $n=30$ stars, we generated synthetic $\{R \sin i\}$ datasets with these parameters by applying different levels of censoring and truncation. 
Figure\,\ref{fig:test_synth_censored} reports some examples in the censoring no-truncation cases.
These tests show that for the Pleiades dataset considered here, where $R \sin i$ truncation does not exceed 0.4, the maximum censoring is around $(v \sin i)_{\rm lim} \sim 7$ km s$^{-1}$, and $\sigma / \mathcal{R} \approx 0.1$, we expect that $\mathcal{R}$ can be recovered with a precision better than 2\% and $\sigma$ with a precision better than 1\%.
We note that censoring at $(v \sin i)_{\rm lim} \apprge 9$ km s$^{-1}$ affects the core of the distribution, making it difficult to recover its original shape.

To evaluate how the accuracy depends on the number of measurements and on the $\sigma/\mathcal{R}$ ratio, we applied the method to groups of 100 $R \sin i$ synthetic distribution realisations, each one with a different number of stars in each bin and different $\sigma/\mathcal{R}$ values.
For brevity, here we compare the results obtained with no-censoring and no-truncation with the worse levels of censoring and truncation in the Pleiades dataset.
Furthermore, we compare the results obtained for $\sigma /\mathcal{R} = 0.1$, which is representative of non-extreme values of the radius dispersion, with those obtained with $\sigma /\mathcal{R} = 0.03$ and 0.3, which are representative of extremely low and extremely high values of the radius dispersion.
We recall that $\sigma$ takes both the intrinsic radius dispersion and the observational uncertainties into account.

\begin{figure*}[ht]
\begin{center}
\includegraphics[width=0.4\textwidth]{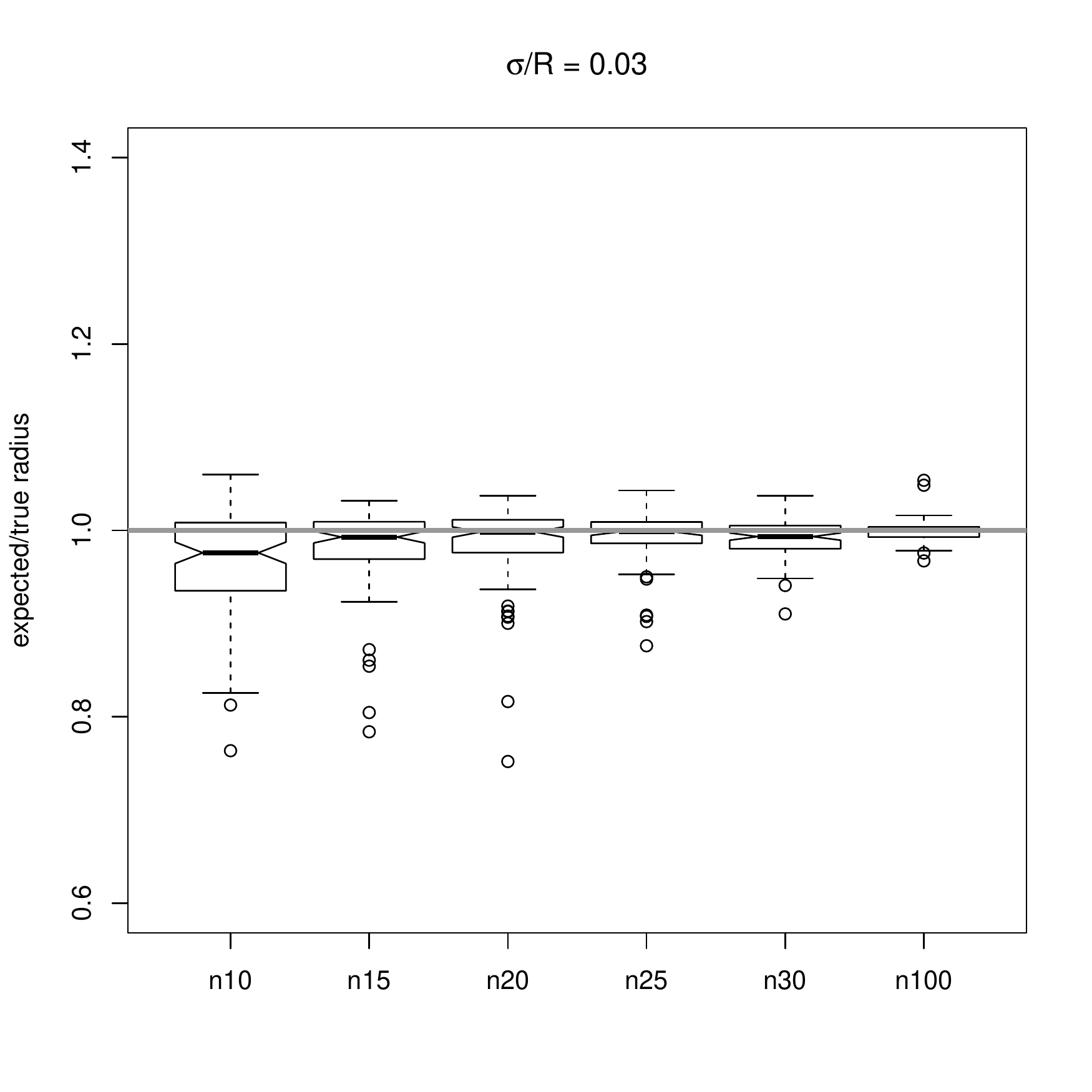}
\includegraphics[width=0.4\textwidth]{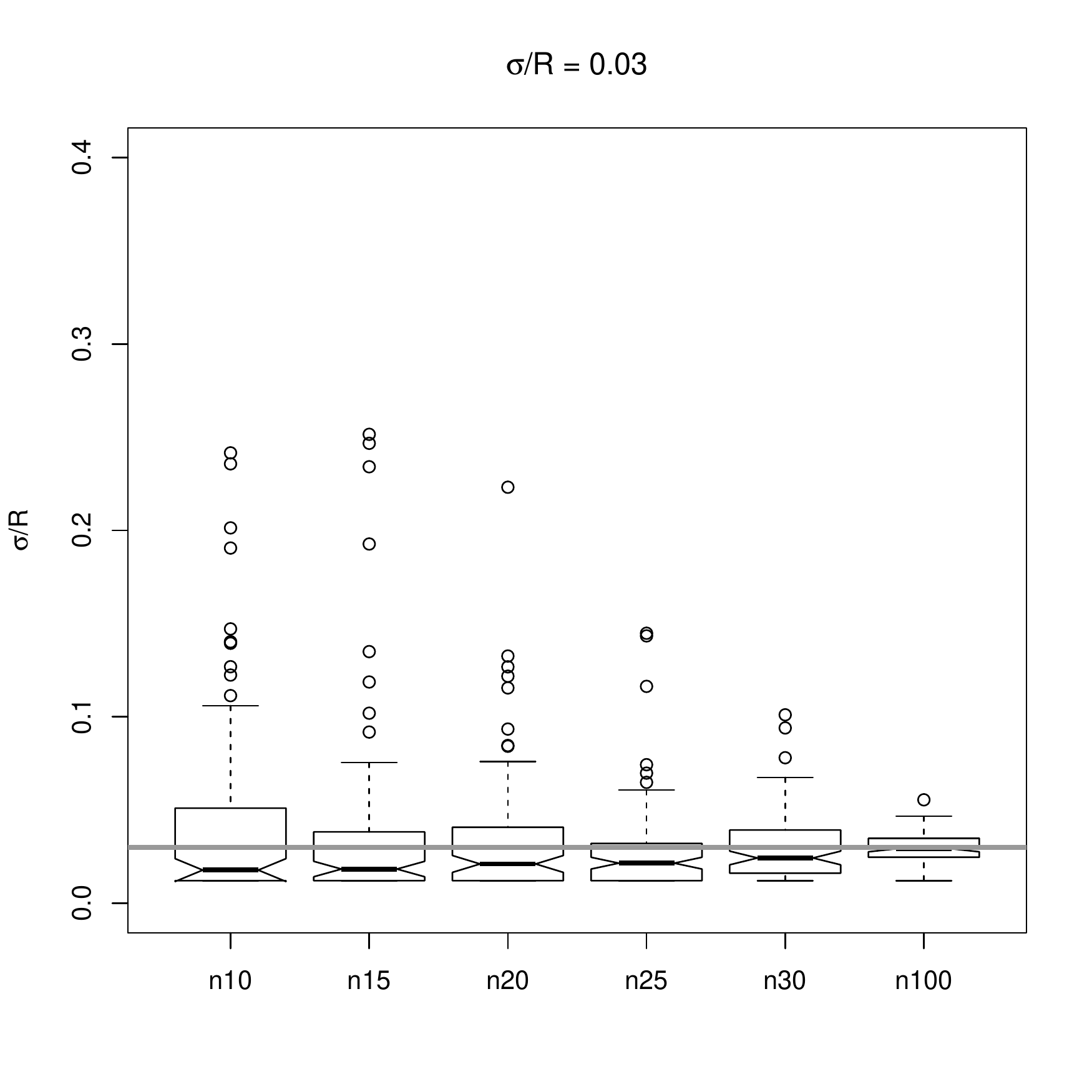}
\includegraphics[width=0.4\textwidth]{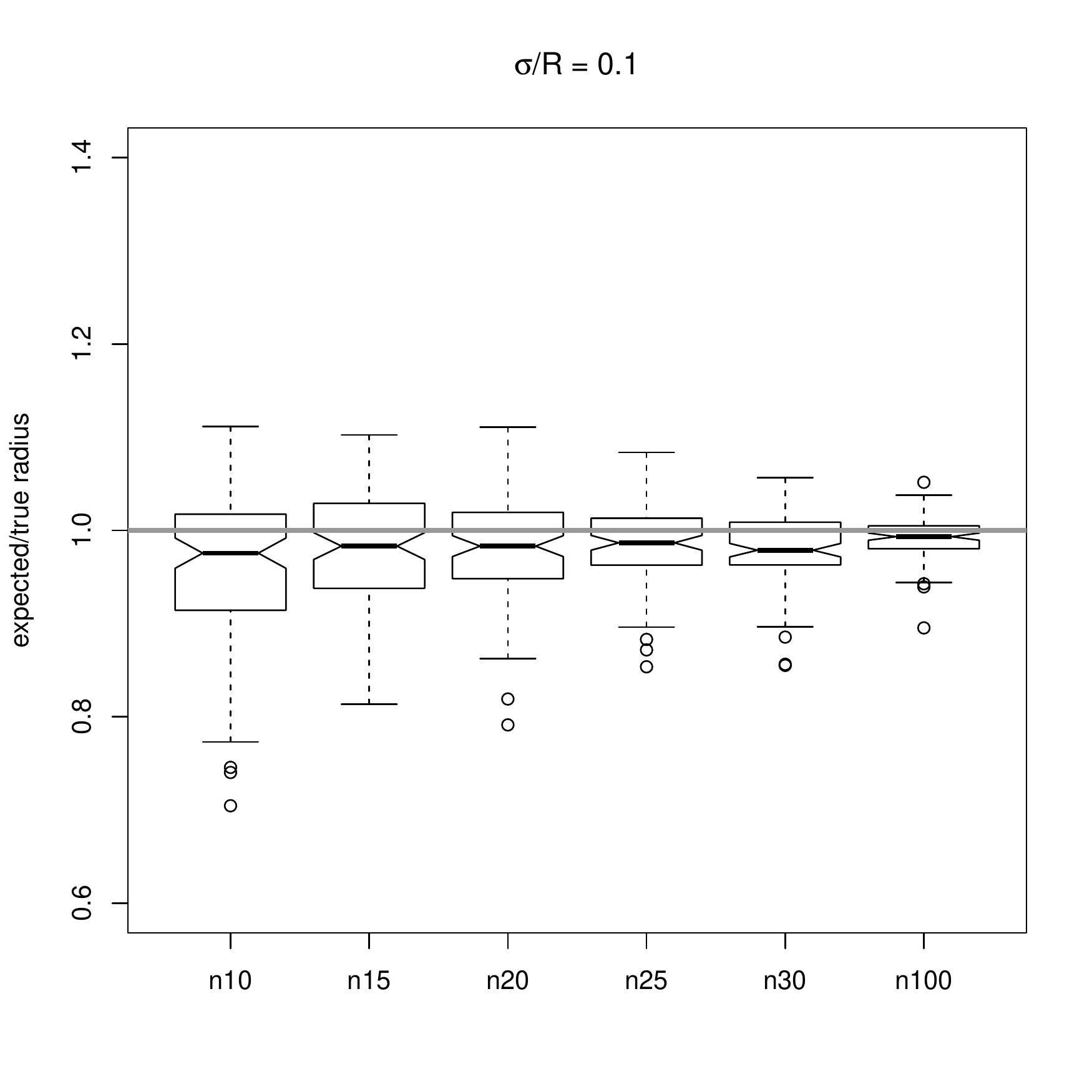}
\includegraphics[width=0.4\textwidth]{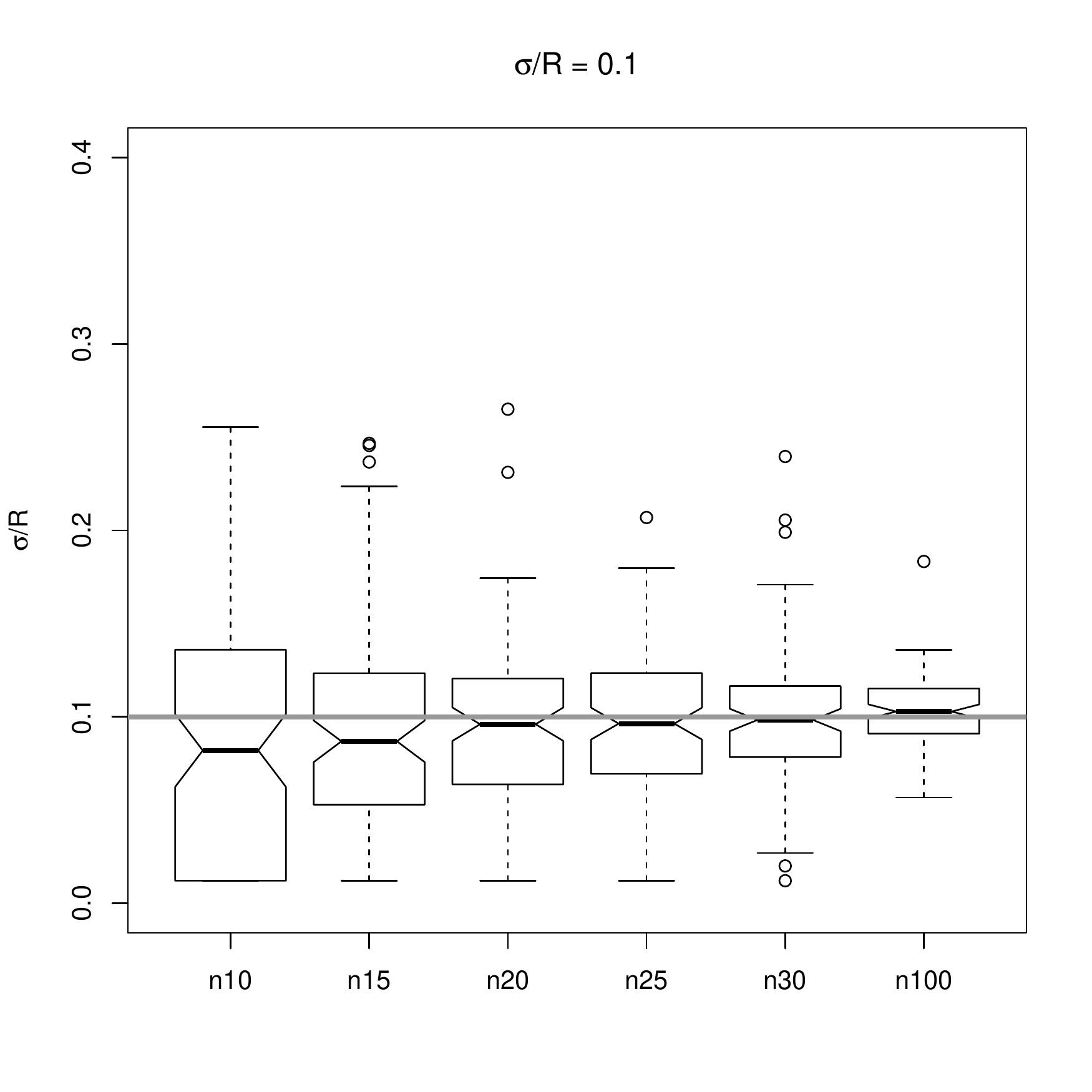}
\includegraphics[width=0.4\textwidth]{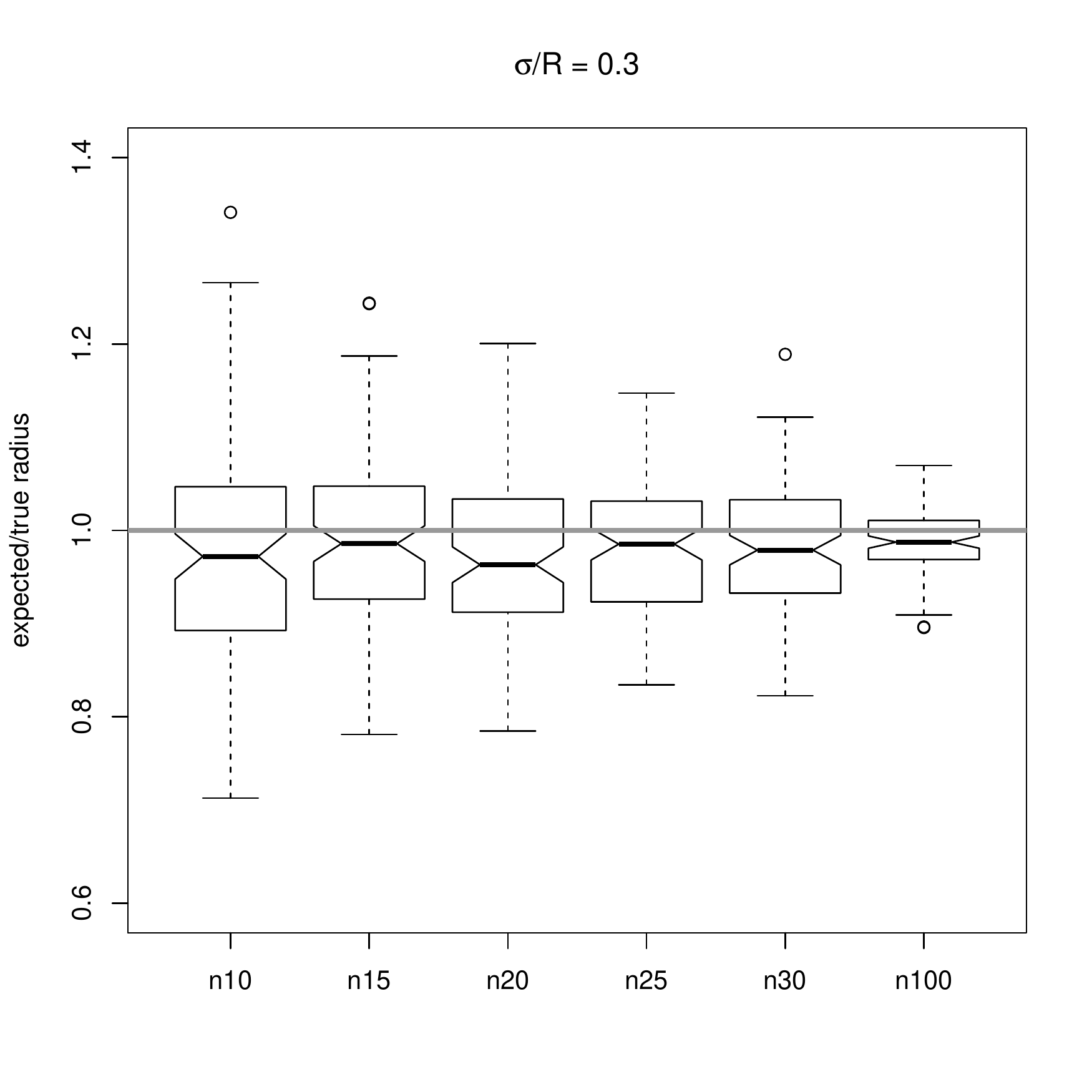}
\includegraphics[width=0.4\textwidth]{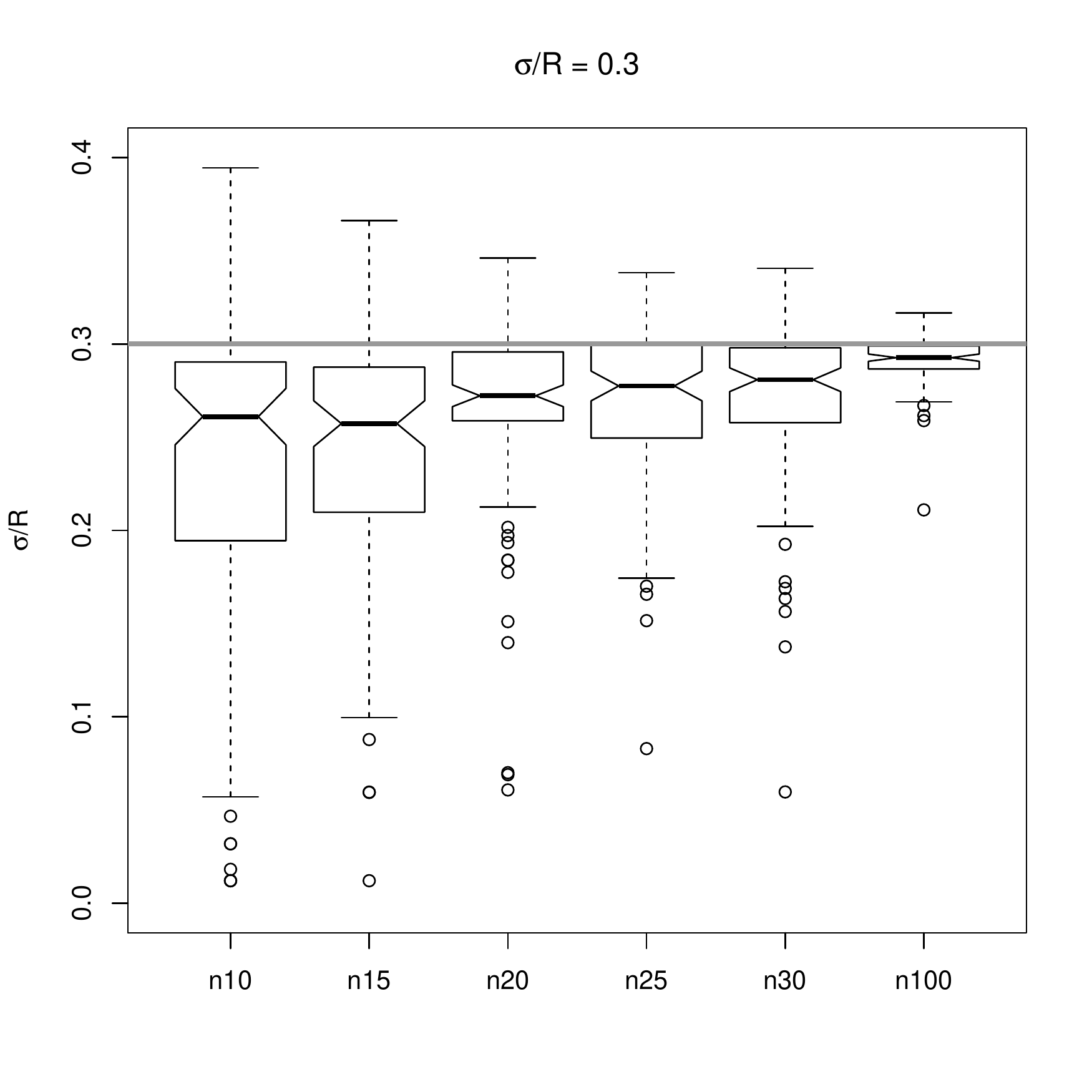}
\caption{Box-and-whisker plots for the reconstructed $\mathcal{R}$ and $\sigma$ from 100 random realisations of the $R \sin i$ distribution with no censoring and no truncation at different $\sigma / \mathcal{R}$ values and different number of measurements per bin.
}
\label{fig:SimuNocensNotrunc}
\end{center}
\end{figure*}

Figure\,\ref{fig:SimuNocensNotrunc} shows the results for no censoring and no truncation.
For $\sigma /\mathcal{R} = 0.1$ the median of the ratio of the expected value to the true value of $\mathcal{R}$ ranges from 0.976 for $n=10$ to 0.993 for $n=100$.
For narrow distributions (e.g. $\sigma /\mathcal{R} = 0.03$) this is essentially unity with $n=15$ or more while for extremely broad distributions (e.g. $\sigma /\mathcal{R} = 0.3$) it amounts to 0.963 in the worst case.
The scatter in the expected-to-true $\mathcal{R}$ ratio decreases with decreasing width of the distribution and with increasing number of observations.
The median of the reconstructed $\sigma$ behaves in a similar way, although its relative accuracy decreases more significantly with increasing width of the distribution.
The intrinsic skewness of the $R \sin i$ distribution leads to a general tendency of underestimating the true $\mathcal{R}$ and $\sigma$ for small $n$ and large  $\sigma / \mathcal{R}$.

\begin{figure*}[ht]
\begin{center}
\includegraphics[width=0.4\textwidth]{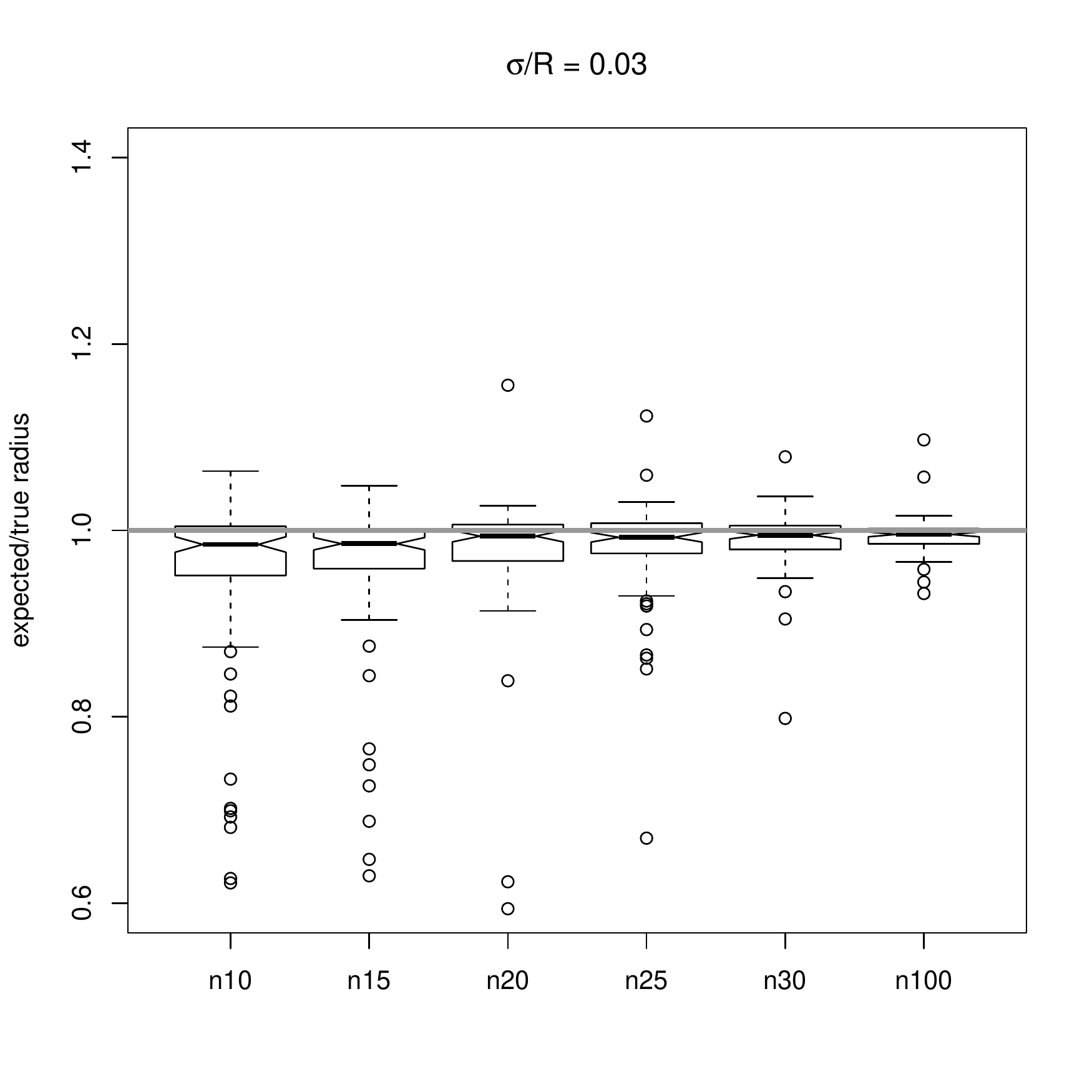}
\includegraphics[width=0.4\textwidth]{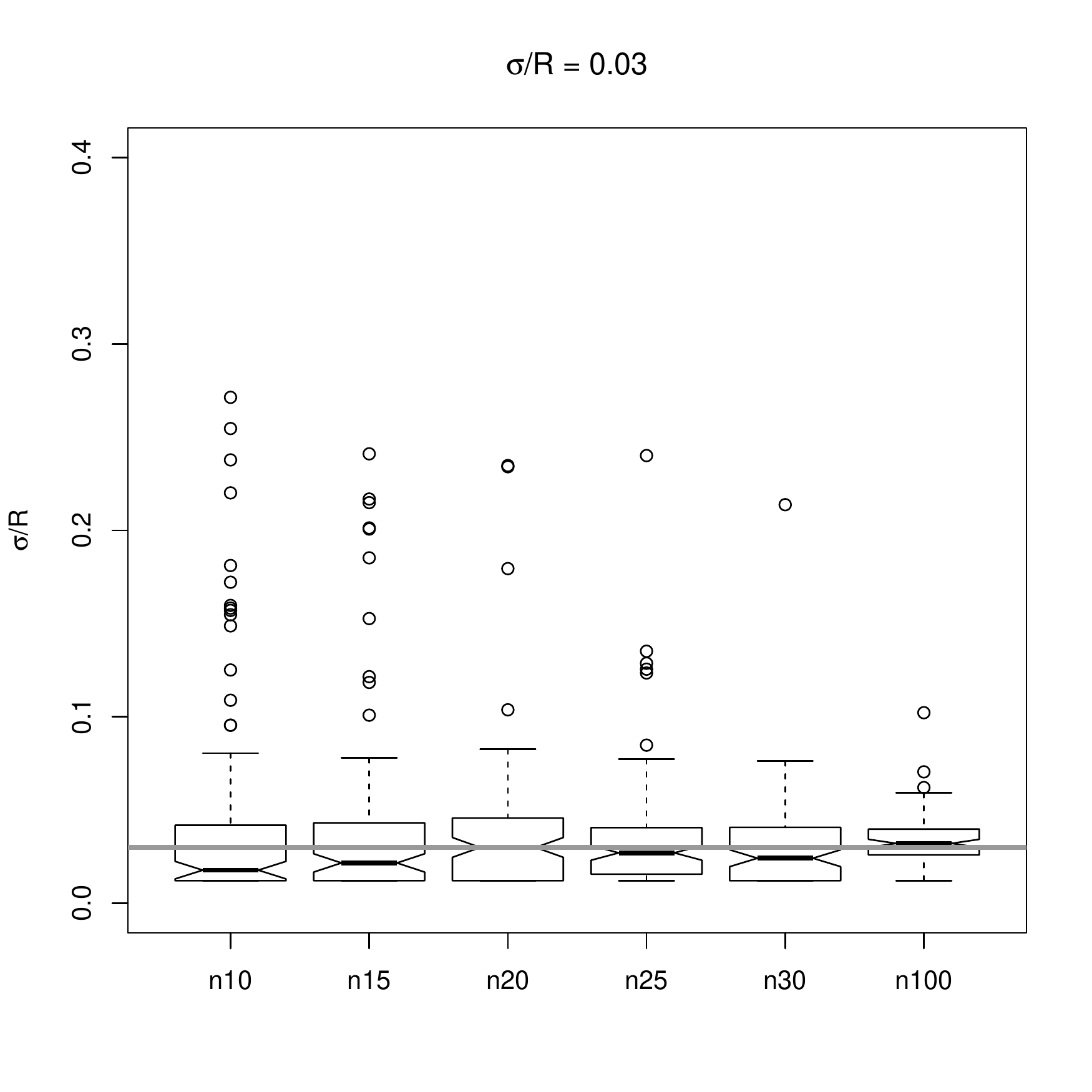}
\includegraphics[width=0.4\textwidth]{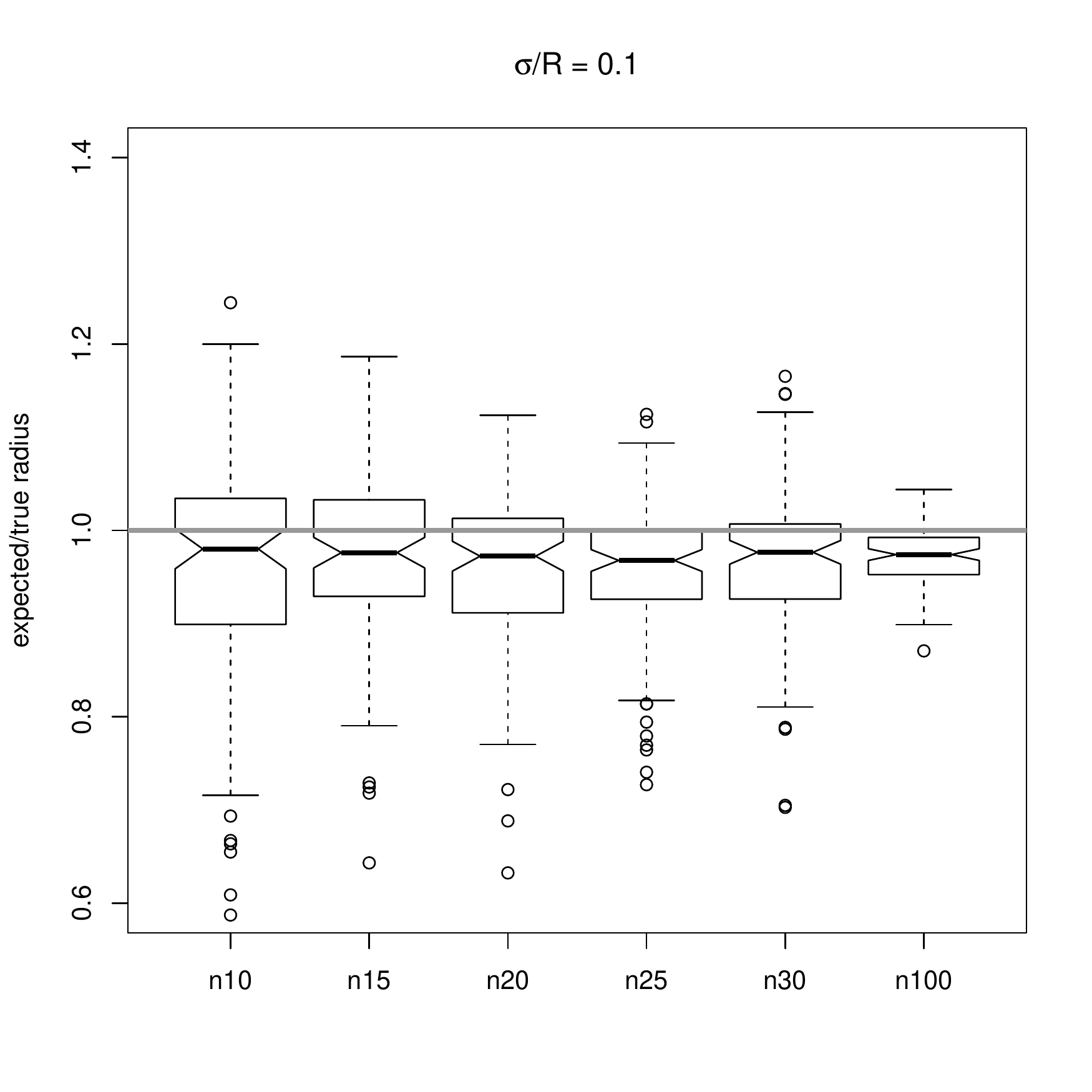}
\includegraphics[width=0.4\textwidth]{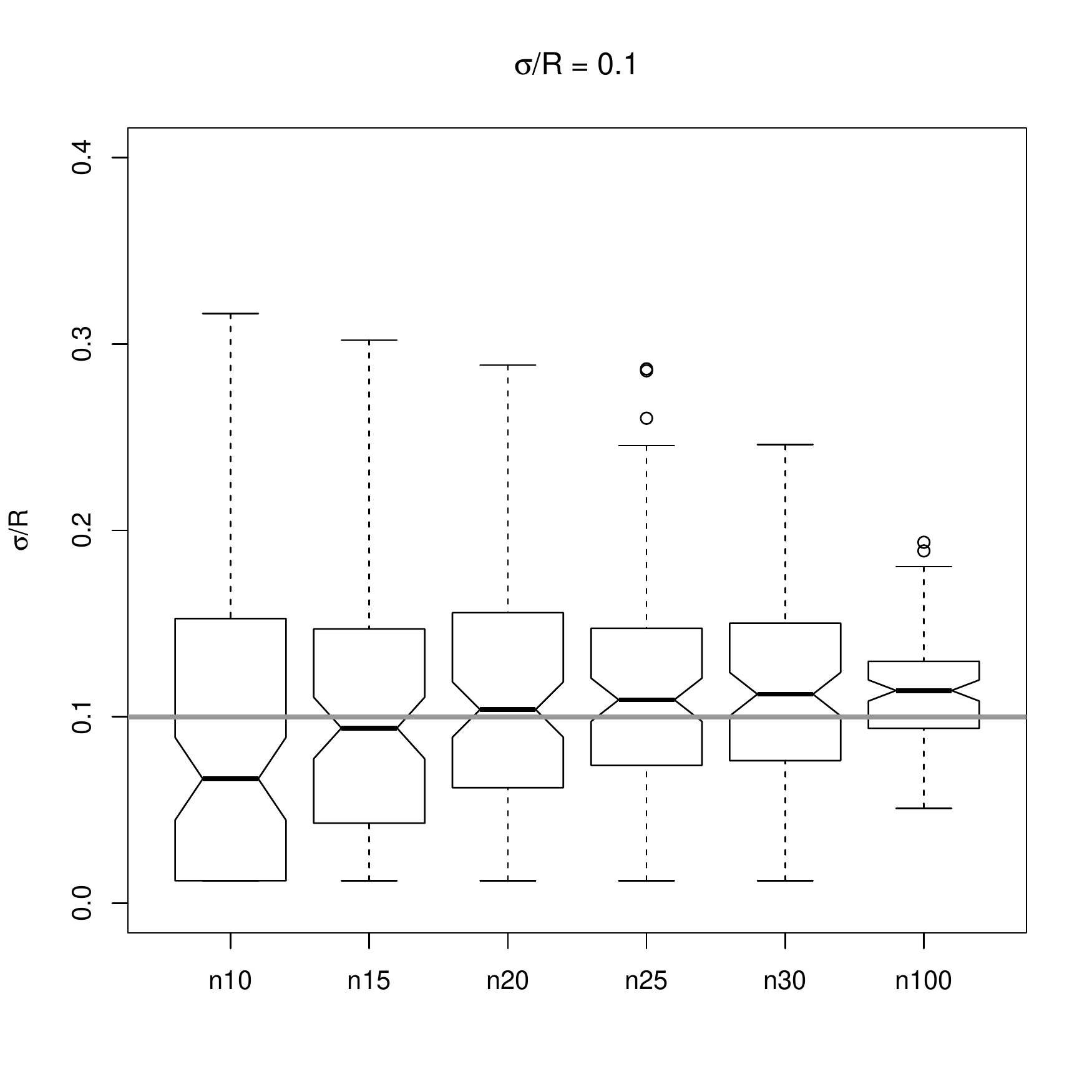}
\includegraphics[width=0.4\textwidth]{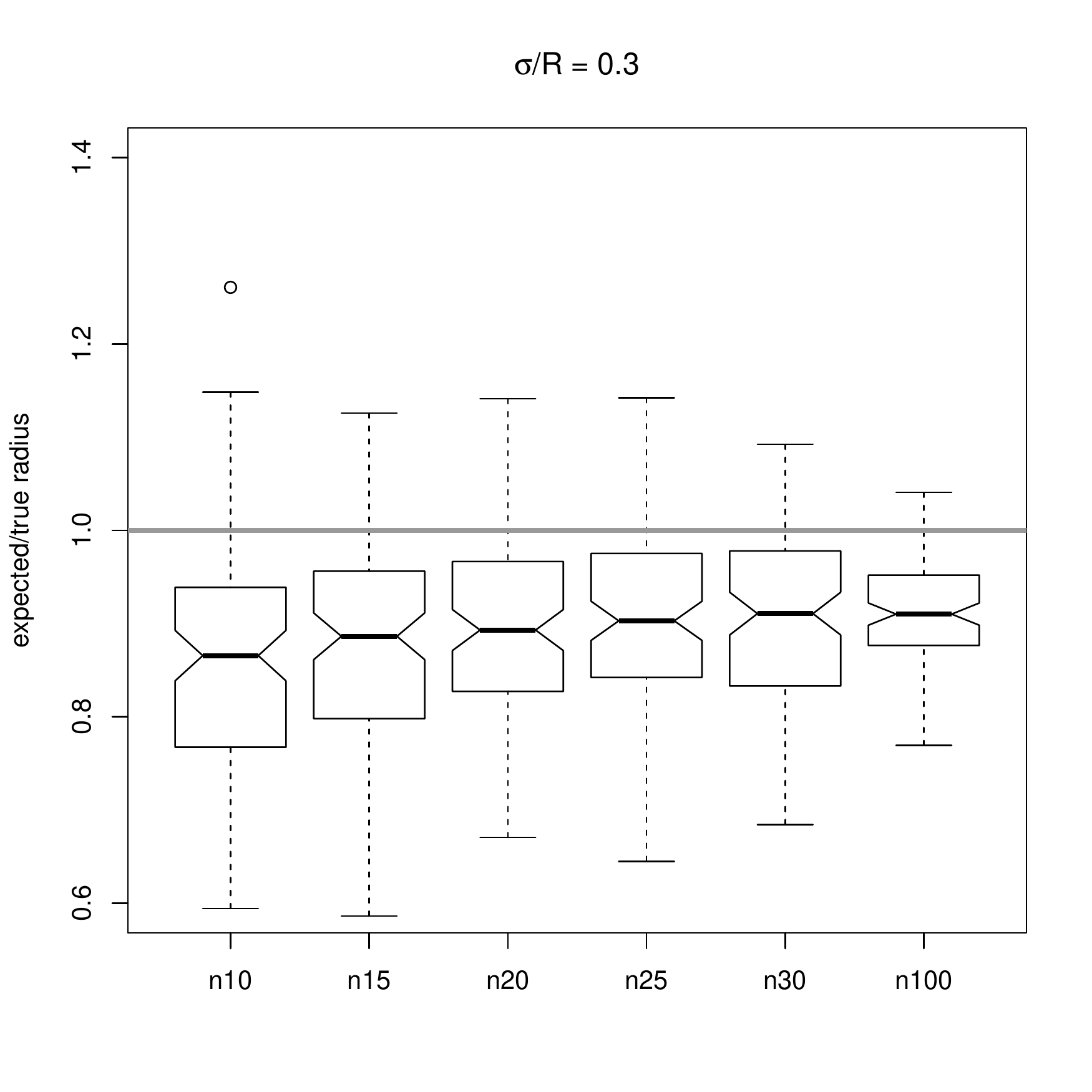}
\includegraphics[width=0.4\textwidth]{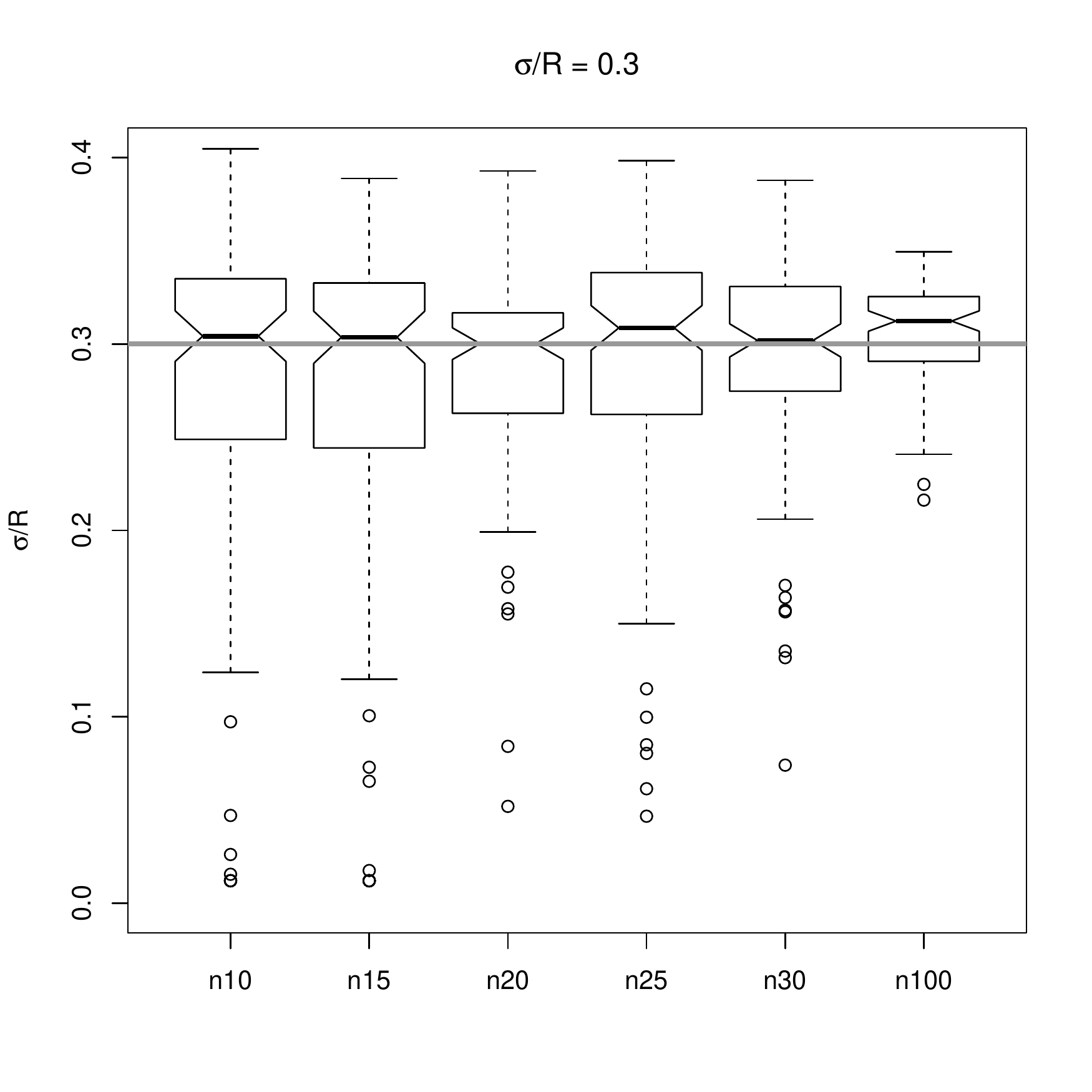}
\caption{Box-and-whisker plots for the reconstructed $\mathcal{R}$ and $\sigma$ from 100 random realisations of the $R \sin i$ distribution with censoring and truncation at different $\sigma / \mathcal{R}$ values and different number of measurements per bin.
}
\label{fig:SimuCensTrunc}
\end{center}
\end{figure*}

The results of the simulation with the worse level of censoring and truncation in the Pleiades dataset are summarised in Fig.\,\ref{fig:SimuCensTrunc}.
For sufficiently narrow distributions the accuracy does not degrade significantly with respect to the no-censoring and no-truncation cases.
Only for large $\sigma / \mathcal{R}$, that is, when censoring and truncation affect the peak of the $R \sin i$ distribution, the median of the ratio of the expected value to true value of $\mathcal{R}$ is significantly below unity.
For the $\sigma / \mathcal{R} = 0.3$ case shown in Fig.\,\ref{fig:SimuCensTrunc}, the radius is underestimated by $\approx$ 10 percent (median) even when the number of measurements is increased to $n=100$.
We note, however, that this latter condition is never met in the Pleiades dataset we analysed in this paper and it is presented here to outline a condition in which it is not possible to recover the average radius reliably.

In summary, the golden rule for evaluating the mean radius is that censoring and truncation must not affect the core of the $R \sin i$ distribution. 
We argue that this is a general requirement, which is not due to a limitation of this particular method, but to the lack of sufficient information when the data cannot define the core of the $R \sin i$ distribution with sufficient
detail.

\end{document}